\documentclass{emulateapj}

\usepackage{amsmath}
\usepackage{amssymb}
\usepackage{graphicx}
\usepackage{color}
\usepackage{natbib}
\usepackage{epsfig}

\def\gtaprx {\lower .1ex\hbox{\rlap{\raise .6ex\hbox{\hskip .3ex
	{\ifmmode{\scriptscriptstyle >}\else
		{$\scriptscriptstyle >$}\fi}}}
	\kern -.4ex{\ifmmode{\scriptscriptstyle \sim}\else
		{$\scriptscriptstyle\sim$}\fi}}}
\def\ltaprx {\lower .1ex\hbox{\rlap{\raise .6ex\hbox{\hskip .3ex
	{\ifmmode{\scriptscriptstyle <}\else
		{$\scriptscriptstyle <$}\fi}}}
	\kern -.4ex{\ifmmode{\scriptscriptstyle \sim}\else
		{$\scriptscriptstyle\sim$}\fi}}}

\newcommand{\cutt}[1]{\textcolor{blue}{}}

\newcommand{\Ms}{{\ensuremath{{M}_{\odot} }}}
\newcommand{\Zs}{\ensuremath{Z_\odot}}

\newcommand{\Ni}{{\ensuremath{^{56}\mathrm{Ni}}}}

\newcommand{\Co}{{\ensuremath{^{56}\mathrm{Co}}}}

\newcommand{\HII}{{\ion{H}{2}}}

\begin{document}

\title{Finding the First Cosmic Explosions. IV. 90 - 140 \Ms\ Pair-Instability Supernovae}

\author{Joseph Smidt\altaffilmark{1}, Daniel J. Whalen\altaffilmark{2}, E. 
Chatzopoulos\altaffilmark{3,4}, Brandon K. Wiggins\altaffilmark{5}, Ke-Jung 
Chen\altaffilmark{6}, Alexandra Kozyreva\altaffilmark{7} 
and Wesley Even\altaffilmark{8}}

\altaffiltext{1}{XTD-IDA, Los Alamos National Laboratory, Los Alamos, NM 87545}

\altaffiltext{2}{Universit\"{a}t Heidelberg, Zentrum f\"{u}r Astronomie, Institut f\"{u}r 
Theoretische Astrophysik, Albert-Ueberle-Str. 2, 69120 Heidelberg, Germany;
dwhalen1999@gmail.com}

\altaffiltext{3}{Department of Astronomy \& Astrophysics, Flash Center for Computational 
Science, University of Chicago, Chicago, IL 60637, USA}

\altaffiltext{4}{Enrico Fermi Fellow}

\altaffiltext{5}{Department of Physics and Astronomy, Brigham Young University, Provo, 
UT  84602}

\altaffiltext{6}{Department of Astronomy \& Astrophysics, University of California, Santa 
Cruz, CA 95064} 

\altaffiltext{7}{Argelander-Institut f{\"u}r Astronomie, Universit{\"a}t Bonn, Auf dem H{\"u}gel 
71, 53121 Bonn, Germany}

\altaffiltext{8}{CCS-2, Los Alamos National Laboratory, Los Alamos, NM 87545}

\begin{abstract}

Population III stars that die as pair-instability supernovae are usually thought to fall in 
the mass range of 140 - 260 \Ms.  But several lines of work have now shown that rotation 
can build up the He cores needed to encounter the pair instability at stellar masses as low 
as 90 \Ms.  Depending on the slope of the initial mass function of Population III stars, there 
could be 4 - 5 times as many stars from 90 - 140 \Ms\ in the primordial universe than in the 
usually accepted range.  We present numerical simulations of the pair-instability explosions 
of such stars performed with the MESA, FLASH and RAGE codes.  We find that they will be 
visible to supernova factories such as Pan-STARRS and LSST in the optical out to $z \sim$ 
1 - 2 and to {\it JWST} and the 30 m-class telescopes in the NIR out to $z \sim$ 7 - 10.  
Such explosions will thus probe the stellar populations of the first galaxies and cosmic star 
formation rates in the era of cosmological reionization.  These supernovae are also easily 
distinguished from more massive pair-instability explosions, underscoring the fact that there 
is far greater variety to the light curves of these events than previously understood.  

\vspace{0.1in}

\end{abstract}


\section{Introduction}

Pair-instability supernovae (PI SNe) are the most energetic thermonuclear explosions 
known, and can be detected near the edge of the observable universe.  They have now 
been studied by several groups for their potential to probe the properties of the first 
stars and galaxies \citep{turk09,stacy10,clark11,hos11,sm11,get11,get12,stacy12,
dw12,glov12,hir13,susa13,get08,jlj09,get10,jeon11,pmb11,wise12,pmb12,jet14}.  They 
can also shed light on the origins of supermassive black holes and early cosmological 
reionization and chemical enrichment \citep{awa09,th09,pm11,wf12,pm12,jlj12a,vol12,
agarw12,jet13,pm13,latif13c,latif13a,schl13,choi13,reis13,jet14,wan04,wet08b,wet10,
awb07,wa08a,mbh03,ss07,bsmith09,chiaki12,ritt12,ss13}.  PI SN candidates such as 
SN 2007bi \citep{gy09,kz14a} and SN 2213 - 1745 \citep{cooke12} have now been 
discovered at $z =$ 0.126 and 2.05, respectively.

These studies have shown that 140 - 260 \Ms\ Population III (Pop III) PI SNe are visible 
in the near infrared (NIR) at $z \gtrsim$ 30 to the \textit{James Webb Space Telescope} 
\citep[\textit{JWST};][]{fwf10,kasen11,jw11,wet12b,wet12a,hum12,pan12a,wet13d,ds13,
ds14} \citep[see also][]{hw02,sc05,ky05,wet08a,chen14a,chen14c}.  They, along with 
Pop III gamma-ray bursts \citep[GRBs; e.g.,][]{wet08c,nak12,met12a,mes13a}, will also 
be visible at $z \sim$ 10 - 20 to the Wide-Field Infrared Survey Telescope (WFIRST) and 
Wide-field Imaging Surveyor for High Redshift (WISH), and at $z <$ 10 to Euclid.  Less 
energetic Pop III SNe will be visible to \textit{JWST} at $z \sim$ 10 - 20, depending on 
explosion type \citep{tomin11,moriya12,tet12,wet12e,wet12c,tet13,wet13c,wet13d,
smidt13a} \citep[see also][for new work on supermassive Pop III SNe]{wet12d,jet13a,
wet13a,wet13b,chen14b}.

This picture changes at higher metallicities.  New explosion models of 150 - 500 \Ms\ PI 
SNe at Large Magellanic Cloud (LMC) and Small Magellanic Cloud (SMC) metallicities 
\citep{wet13e} have light curves that are quite different from those of zero-metallicity 
explosions, for two reasons.  First, stars at these metallicities lose most of their mass to 
strong winds or outbursts that form structures around the star that can either quench or 
brighten emission from the shock. Mass loss also reduces the star to a compact He core 
by the time it dies, with $\sim$ 1\% of the original radius of the star.  \Ni\ yields and 
radiation diffusion timescales out of the ejecta are very different for explosions of bare He 
cores than for stars that retain their H envelope. Such explosions can therefore either be 
dim events that can only be seen in the local universe or superluminous events that are 
visible out to high redshifts. These new studies underscore the fact that there is far more 
variety to PI SN light curves than previously imagined.

How does rotation alter the energies and luminosities of Pop III PI SNe?  It is now known 
that rotation can build up He cores massive enough to encounter the pair instability at 
stellar masses well below 140 \Ms.  \citet{cw12} (hereafter CW12) have shown that 90 - 
135 \Ms\ Pop III stars can explode as PI SNe if they if they are born with rotation rates at 
50\% of the breakup velocity.  They die as compact He cores because rotational mixing 
dredges heavier elements up to the outer layers of the star and drives mass loss.  Their 
compact geometries guarantee that their light curves will be different from those of more 
massive stars that have retained their envelopes.  Rotation can also induce bulging in the 
equators and flattening in the poles of such stars, which could introduce an azimuthal 
dependence to their light curves \citep{cwc13}.  

How rotation affects the luminosities of ancient PI SNe is important because recent studies 
suggest that some massive Pop III stars may have been born with angular velocities close 
to the breakup limit \citep{get11,stacy13}.  Rotation may also have enabled much higher 
numbers of PI SNe at high redshift because 90 - 135 \Ms\ Pop III stars could be 4 - 5 times 
more numerous as those previously studied, depending on their initial mass function (IMF).  
Given their compact explosion geometries and variety of energies and \Ni\ yields, can these 
events also probe the properties of the first stars?  We have now modeled light curves and 
spectra for 90 - 135 \Ms\ PI SNe with the Los Alamos RAGE and SPECTRUM codes.  In 
Section 2 we review our stellar evolution and initial explosion models along with our RAGE 
and SPECTRUM simulations.  The blast profiles are examined in Section 3, and NIR light 
curves and detection limits for these explosions as a function of redshift are presented in 
Section 4.  We conclude in Section 5.

\section{Numerical Models}

We calculate light curves and spectra for PI SNe in five stages.  First, the stars are 
evolved from the zero-age main sequence (ZAMS) in the MESA code up to the onset of 
the PI.  At this point the models are mapped into the FLASH code and exploded.  When 
nuclear burning is complete, typically within a few tens of seconds, we port the profile for 
the shock, the surrounding star, and the ambient wind into the RAGE code and evolve 
the SN out to 3 yr.  We then post process our RAGE profiles with the SPECTRUM code 
to construct light curves and spectra. Finally, these spectra are cosmologically redshifted 
and dimmed to obtain NIR light curves in the observer frame. 

\subsection{MESA / FLASH Simulations}

The progenitor stars considered here are those from CW12, who studied the effects of 
rotation on the minimum masses of both pair-pulsational (PP) and PI SNe.  These stars 
were evolved in the one-dimensional (1D) Lagrangian stellar evolution code MESA \citep{
paxt11,paxt13}, which includes a parametrized treatment of rotation and magnetic fields.  
How rotation induces mixing and angular momentum transport in these stars is discussed 
in detail in CW12.  Although mass loss for highly-evolved massive Pop III stars is not 
observationally constrained, it must be included because it can affect rotation by allowing 
the star to shed excess angular momentum over time as it evolves. In lieu of actual 
observations, we adopt the prescription of \citet{dj88} and \cite{Vink01} for mass loss from 
the stars in our models.  

We modify this loss rate, which the star would have even if it was stationary, to account for 
rotation according to the method of \citet{hlw00}: \vspace{0.1in}
\begin{equation}
\dot{m} = \dot{m}_{\rm no-rot}/(1-\Omega/\Omega_{c})^{0.43}, \vspace{0.1in}
\end{equation}
where $\dot{m}_{\rm no-rot}$ is the mass-loss rate from \citet{dj88} and \citet{Vink01} and 
$\Omega$ is the surface angular velocity at the stellar equator.  When $\Omega/\Omega
_{c} =$~1 $\dot{m}$ diverges, so the mass loss timescale in MESA is limited to the thermal 
timescale of the star, $\tau_{KH}$: $\dot{m} = {\rm min} (\dot{m}(\Omega),f m/\tau_{KH})$, 
where $f$ is an efficiency factor taken to be 0.3 \citep{ywl10}.  To be consistent with \citet{
get11} and \citet{stacy13}, we consider only stars that rotate at 50\% of the breakup velocity 
at ZAMS. CW12 found that the minimum mass for a zero-metallicity PI SN progenitor at this 
initial rotation rate is $\sim$~85~\Ms \citep[see also][]{yoon12}. Our grid of models therefore 
ranges from 90 - 140 \Ms, in 5 \Ms\ increments.  The stars all die as compact cores ($r_{f} 
\sim$ 10$^{10}$ - 10$^{11}$ cm) that are H and, sometimes, He deficient.

\begin{figure}
\includegraphics[width=0.4\textwidth, angle =270 ]{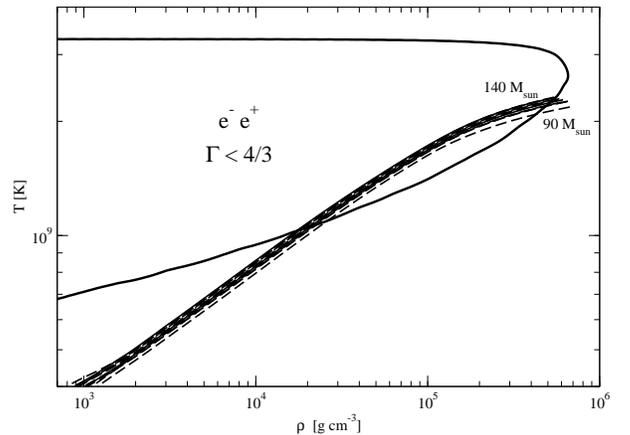}
\caption{Density and temperature structures of the stars prior to explosion.  The thick black 
curve encloses the region where HELM EOS implies that $\Gamma_{ad} <$ 4/3, the PI 
regime.  The dashed curves denote the structure of each star from 90 - 140 \Ms.} \vspace{
0.1in}
\label{fig:rhot}
\end{figure}

In MESA, we adopt the Schwarzschild criterion for convection with $\alpha_{\rm MLT} =$ 
2, the \citet{ts00} ``Helmholtz'' equation of state (HELM EOS), which includes contributions 
from e$^{-}$e$^{+}$ pairs, and the ``approx21'' nuclear reaction network \citep{tim99}, 
which has the $\alpha$-chain elements and the intermediate elements linking them through 
$(\alpha, p)(p,\gamma)$ reactions from neutrons and protons all the way up to \Ni\ (mass 
numbers $A$ from 1 to 56). The number of radial zones in the models was 800 -- 1,200 
(the ``mesh\_delta\_coeff" variable in MESA was set to 0.75 -- 0.95).  The stars are evolved 
from the ZAMS until the CO core encounters the PI and the adiabatic index, $\Gamma_{ad}
$, drops below 4/3 and triggers collapse.  Collapse usually proceeds by the time the core 
reaches $^{20}$Ne exhaustion ($X_{Ne,c} \leq$ 0.01, where $X_{Ne,c}$ is the Ne mass 
fraction in the central zone of the model) but before the start of $^{16}$O burning.  At this 
point we halt MESA.  The final properties of the stars are listed in Table~1, and their density 
and temperature structures prior to explosion are shown in Figure~\ref{fig:rhot}. The interiors
of the cores all clearly cross into the PI regime at the end of their lives.  The energies of the
explosions range from $\sim$ 10 - 90 foe, where 1 foe $=$ 10$^{51}$ erg.

The MESA profiles are then conservatively mapped onto a 1D radial mesh in the FLASH 4.0 
adaptive mesh refinement (AMR) code \citep{Fryx00,dub09}.  Since the stars are mapped 
from a Lagrangian grid in mass to an Eulerian mesh in space, care was taken to conserve 
total mass and energy. The resolution we chose in FLASH was higher than that in the MESA 
models, $\sim$ 10$^{5}$ -- 10$^{6}$ cm. FLASH was run with the HELM EOS and the richest 
nuclear reaction network in the code, ``approx19''.  The omission of two neutron rich isotopes, 
$^{56}$Fe and $^{56}$Cr, and relevant neutronization reactions has little effect in these 
simulations because PI SN progenitors never evolve past C/O burning as do core-collapse 
(CC) SN progenitors.

The angular velocities in the MESA models are set to zero in FLASH because \citet{cwc13} 
found that for a given CO core mass (and all else being equal), only extreme rotation can 
change PI SN energies and \Ni\ production.  Our models exhibit only modest degrees of 
rotation in the core when they reach the pair-instability regime ($\Omega/\Omega_{c,core}
$~$\sim$ 0.02 -- 0.06).  We therefore only need to consider the effects of rotation on the 
structure of the star when evaluating its impact on light curves and spectra.  We evolve the 
shock out just below the surface of the star.  Terminating FLASH at this stage ensures that 
no photons from the shock have broken out of the surface of the star.  Radiation transport 
is not required in this calculation because the mean free paths of the photons in the star are 
so short that they are simply advected along by fluid flows, but we include their contribution 
to the EOS.

\setcounter{table}{0}
\begin{deluxetable}{lcccccc}
\tabletypesize{\tiny}
\tablewidth{0pt}
\tablecaption{PI SN progenitor properties.}
\tablehead{
\colhead {$M_{\star}$} &
\colhead {$r_{f}$~(cm)} &
\colhead {$M_{CO}^{\dagger}$} &
\colhead {$M_{Ni}$} &
\colhead {$E_{ex}$~(erg)} 
 &\\}

\startdata
\hline
          &                    &              &              &                 \\
90      &     3.9e10    &    59.3   &   0.14    &   9.9e51   \\  
95      &     5.6e10    &    63.5   &   0.34    &   1.1e52   \\  
100    &     5.6e10    &    65.6   &   0.44    &   1.2e52   \\  
105    &     4.7e10    &    69.1   &   0.90    &   2.8e52   \\   
110    &     5.7e10    &    70.4   &   1.14    &   3.9e52   \\   
%115    &     8.3e10    &    70.5   &   1.07    &   3.1e52   \\   	   	 
120    &     6.6e10    &    72.6   &   1.57    &   4.3e52   \\   
125    &     8.0e10    &    76.8   &   3.26    &   5.0e52   \\   
130    &     1.1e11    &    77.7   &   3.87    &   5.2e52   \\    
135    &     1.8e11    &    79.8   &   4.52    &   6.2e52   \\    
140    &     7.7e10    &    83.7   &   7.30    &   8.0e52   \\  
\enddata 
\tablecomments{All masses are in \Ms. $^{\dagger}$ $M_{CO}$ is the mass of the 
carbon-oxygen core defined within the radius where $X_{C}+X_{O} >$~0.5.}
\end{deluxetable}

\subsection{RAGE}

The explosion is evolved from breakout from the surface of the star out to 3 yr with the Los 
Alamos code RAGE \citep{rage,fet12}. RAGE is an AMR radiation hydrodynamics code with 
grey or multigroup flux-limited diffusion and a second-order conservative Godunov hydro 
scheme.  RAGE uses Los Alamos OPLIB 
opacities\footnote{http://aphysics2/www.t4.lanl.gov/cgi-bin/opacity/tops.pl} \citep{oplib} and 
includes multispecies advection and 2-temperature (2T) radiation transport, in which matter 
and radiation temperatures, although coupled, are evolved separately.  We include the self
gravity of the ejecta and point mass gravity for any material that falls back to the center of
the grid.  We evolve mass fractions for 15 elements:  H, He, C, N, O, Ne, Mg, Si, S, Ar, Ca, 
Ti, Cr, Fe and Ni.  

\subsubsection{Model Setup}

\begin{figure*}
\plottwo{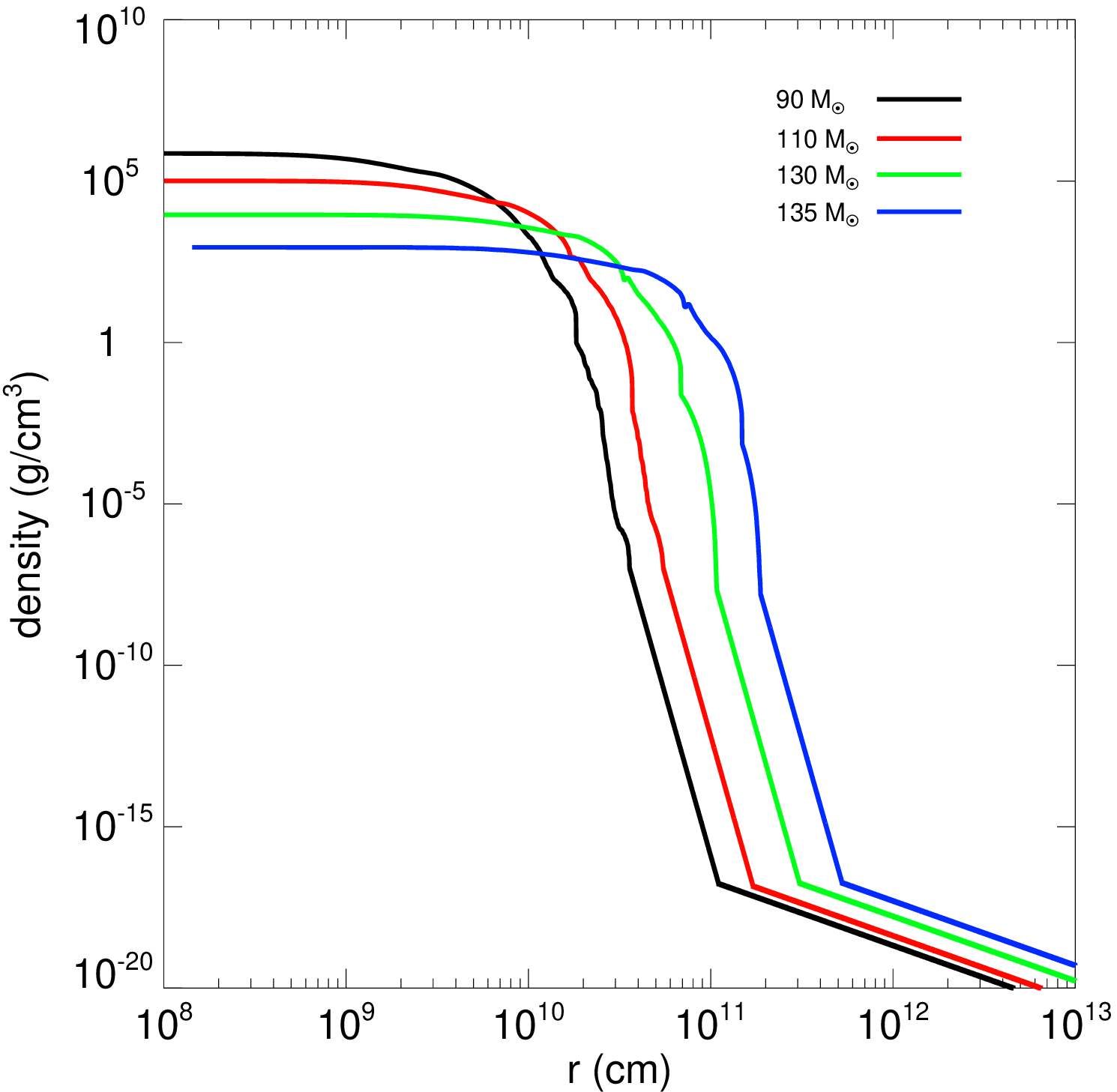}{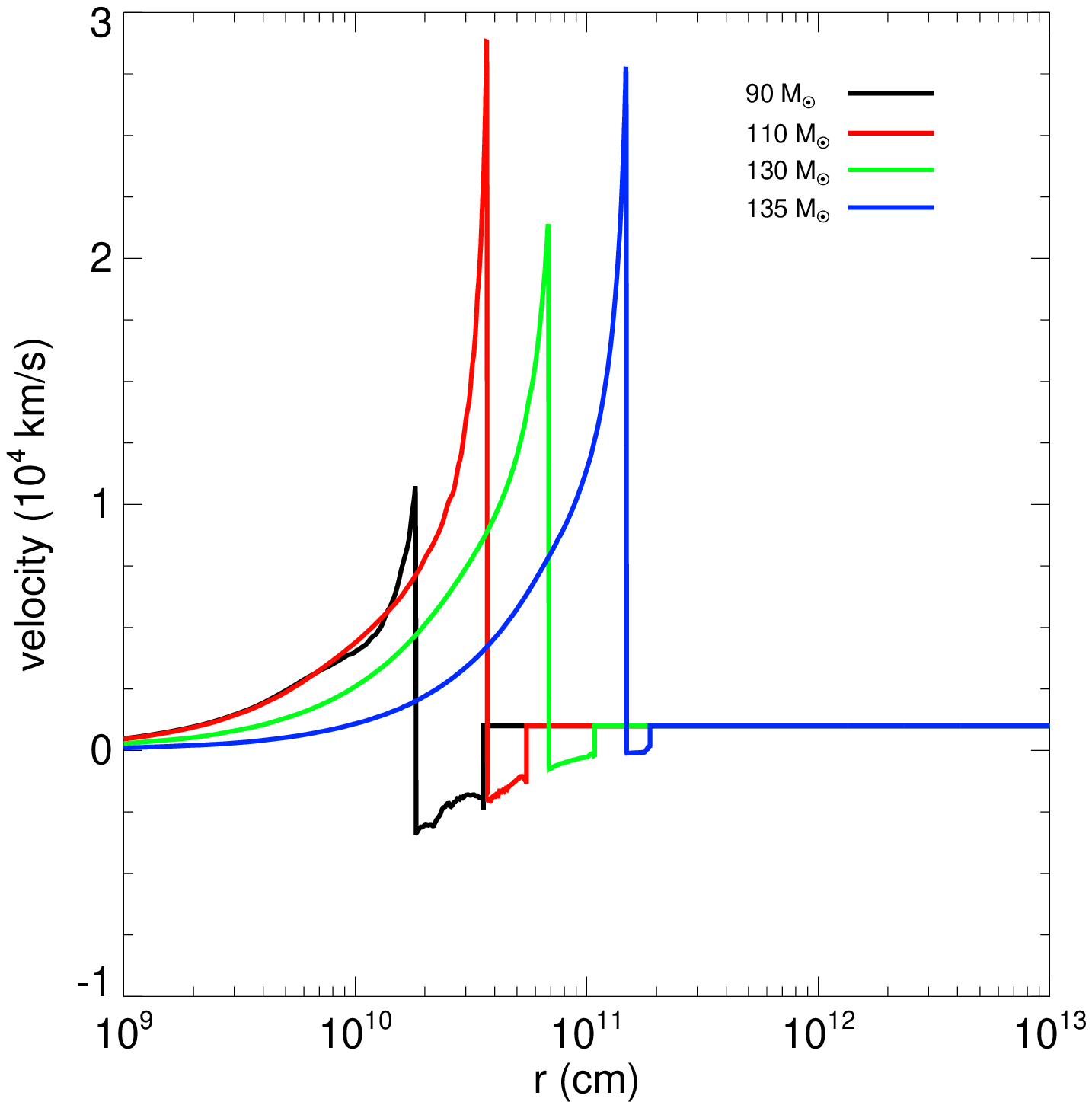}
\caption{RAGE explosion profiles for the 90, 110, 130 and 135 \Ms\ PI SNe.  Left:  densities. 
Right:  velocities} \vspace{0.1in}
\label{fig:setup}
\end{figure*} 

Our 1D spherical coordinate root grid has 100,000 uniform zones with an initial resolution 
that varies from 4 $\times$ 10$^{5}$ cm to 3 $\times$ 10$^{6}$ cm.  We set outflow and 
reflecting boundary conditions on the fluid and radiation flows at the inner boundary of the 
mesh (which is at 0 cm), respectively.  Outflow conditions are set on the gas and radiation 
at the outer boundary. Up to 2 levels of refinement are applied to the initial interpolation of 
the profiles onto the setup grid and then during the simulation.  We initialize radiation 
energy densities in RAGE from the temperatures in the FLASH profiles:
\vspace{0.05in}
\begin{equation}
e_{rad} = aT^4, \vspace{0.05in}
\end{equation}
where $a =$ 7.564 $\times$ 10$^{-15}$ erg cm$^{-3}$ K$^{-4}$ is the radiation constant and 
$T$ is the gas temperature.  We also construct the specific internal energy from $T$: 
\vspace{0.05in}
\begin{equation}
e_{gas} = C_VT, \vspace{0.05in}
\end{equation}
where $C_V = $ 1.2472 $\times$ 10$^{8}$ erg gm$^{-1}$ K$^{-1}$ is the specific heat of the 
gas.  

Our choice of mesh places the shock about a quarter of the way across the grid at launch. 
To accommodate the expansion of the ejecta and speed up the simulation, we resize the 
grid by a factor of 2.5 every 10$^6$ time steps or when the leading edge of the radiation 
front has crossed 90\% of the grid, whichever happens first.  The time step on which the 
new series initially evolves scales approximately as the ratio of the new and old resolutions.  
We join a simple low-mass wind profile to the surface of the star:
\vspace{0.05in}
\begin{equation}
\rho_\mathrm{w}(r) = \frac{\dot{m}}{4 \pi r^2 v_\mathrm{w}}, \vspace{0.05in}
\end{equation}
where $\dot{m}$ is the mass loss rate of the wind and $v_\mathrm{w}$ is its speed.  We 
take $v_\mathrm{w}$ to be 1000 km s$^{-1}$ and the H and He mass fractions in the wind 
to be 76\% and 24\% for simplicity.  The abrupt density drop between the star and wind is 
bridged by an $r^{-20}$ density gradient to avoid numerical instabilities at shock breakout.  
We chose $\dot{m}$ so that $\rho_\mathrm{w} \sim$ 2 $\times$ 10$^{-18}$ g cm$^{-3}$ 
at the bottom of the density bridge, so that it is optically thin there but still dense enough to 
prevent numerical instabilities in the radiation solution.  The wind profile is continued 
outward until its density falls to that of the \HII\ region of the star \citep[e.g.,][]{wan04}. It is 
then replaced by the \HII\ region, which is assumed to have a uniform density $n = $ 0.1 
cm$^{-3}$ and mass fractions of 76\% H and 24\% He.  We show density and velocity 
profiles for a few of our models in Figure~\ref{fig:setup}.       

\subsection{SPECTRUM} 

We calculate a spectrum from a RAGE profile by mapping its densities, temperatures, 
mass fractions and velocities onto a 2D grid in $r$ and $\mu =$ cos $\theta$ in the Los 
Alamos SPECTRUM code.  SPECTRUM directly sums the luminosity of each fluid element 
in the discretized profile to obtain the total flux escaping the ejecta along the line of sight at 
every wavelength. Our method, which is described in detail in \citet{fet12}, includes Doppler 
shifts and time dilation due to the relativistic expansion of the ejecta and the intensities of 
emission lines.  SPECTRUM also accounts for the attenuation of flux along the line of sight, 
capturing both limb darkening and absorption lines imprinted on the flux by intervening 
material in the ejecta and wind.  Each spectrum has 14899 energies.

Velocities, densities, mass fractions and radiation temperatures are extracted from every 
level of the AMR hierarchy in RAGE and sequentially ordered by radius.  Because of 
limitations on machine memory and time, only a subset of this data is mapped into 
SPECTRUM.  We first determine the position of the radiation front, which is taken to be 
where $aT^4$ rises above 10$^{-4}$ erg/cm$^3$.  Next, we find the radius of the $\tau = 
$ 40 surface by integrating the optical depth due to Thomson scattering in from the outer 
edge of the grid, taking $\kappa_{Th}$ to be 0.288 for H and He gas at the mass fractions 
in the wind \citep[see Section 2.4 of][]{wet12c}.  This is the greatest depth from which most 
of the photons can escape the ejecta.

The extracted fluid variables are then interpolated onto the SPECTRUM grid.  The region 
from the center of the grid to the $\tau =$ 40 surface is divided into 800 uniform zones in 
log $r$ and the region from the $\tau =$ 40 surface to the radiation front is partitioned into 
6200 uniform zones in $r$. Five hundred uniform zones in log $r$ are placed between the 
front and the outer edge of the grid, for a total of 7500 radial bins.  The variables in these 
new radial bins are mass averaged to capture sharp features from the RAGE profile.  The 
mesh is uniformly divided into 160 bins in $\mu$ from -1 to 1.  Its inner radial boundary is 
the same as for the RAGE grid and its outer boundary is 10$^{18}$ cm.  The SPECTRUM 
grid fully resolves regions of the ejecta from which photons can escape and only lightly 
samples those from which most cannot.
  
\section{Explosion Profiles}

\begin{figure*}
\begin{center}
\begin{tabular}{ccc}
\epsfig{file=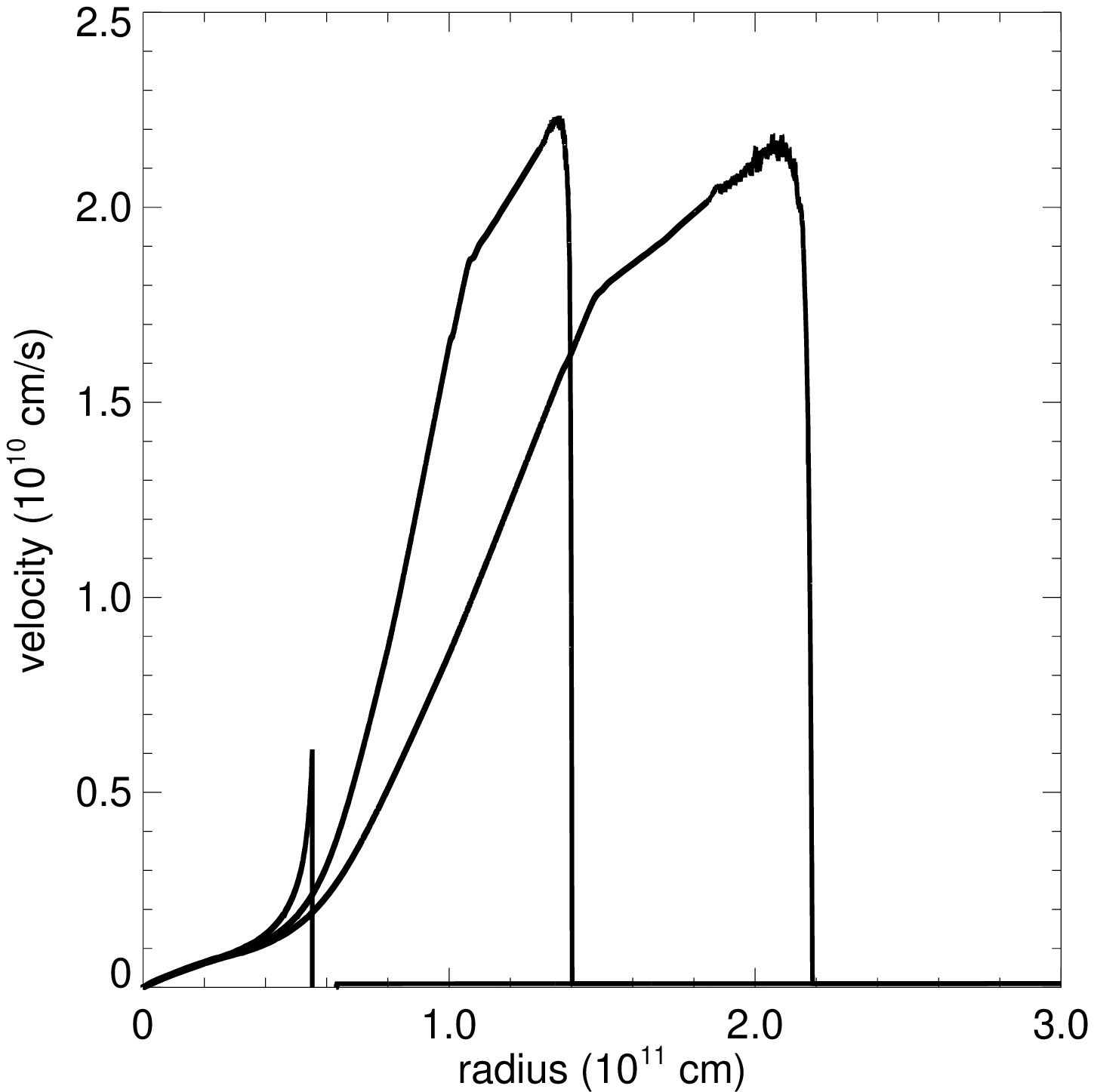,width=0.3\linewidth,clip=}  &
\epsfig{file=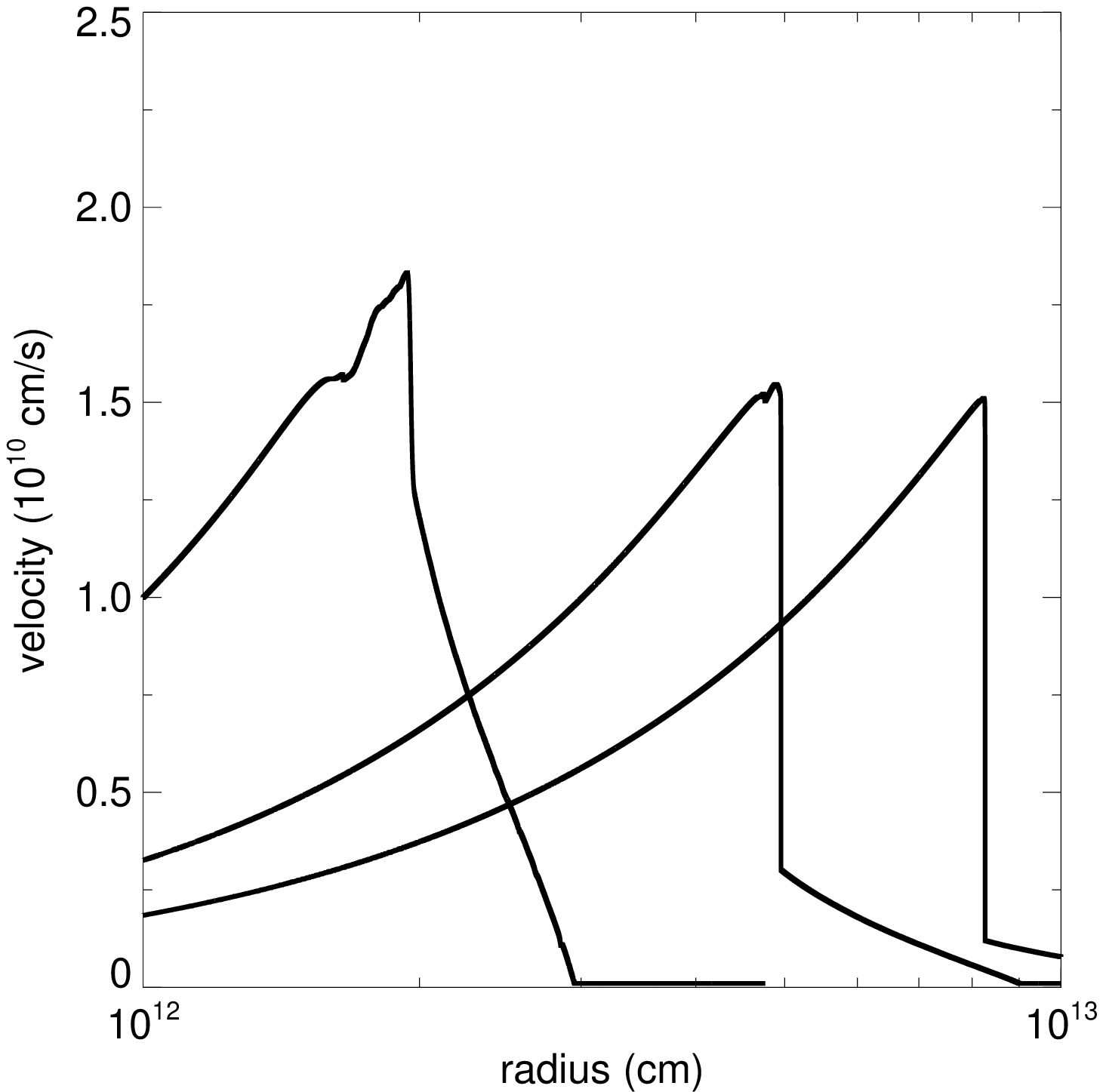,width=0.3\linewidth,clip=}  &
\epsfig{file=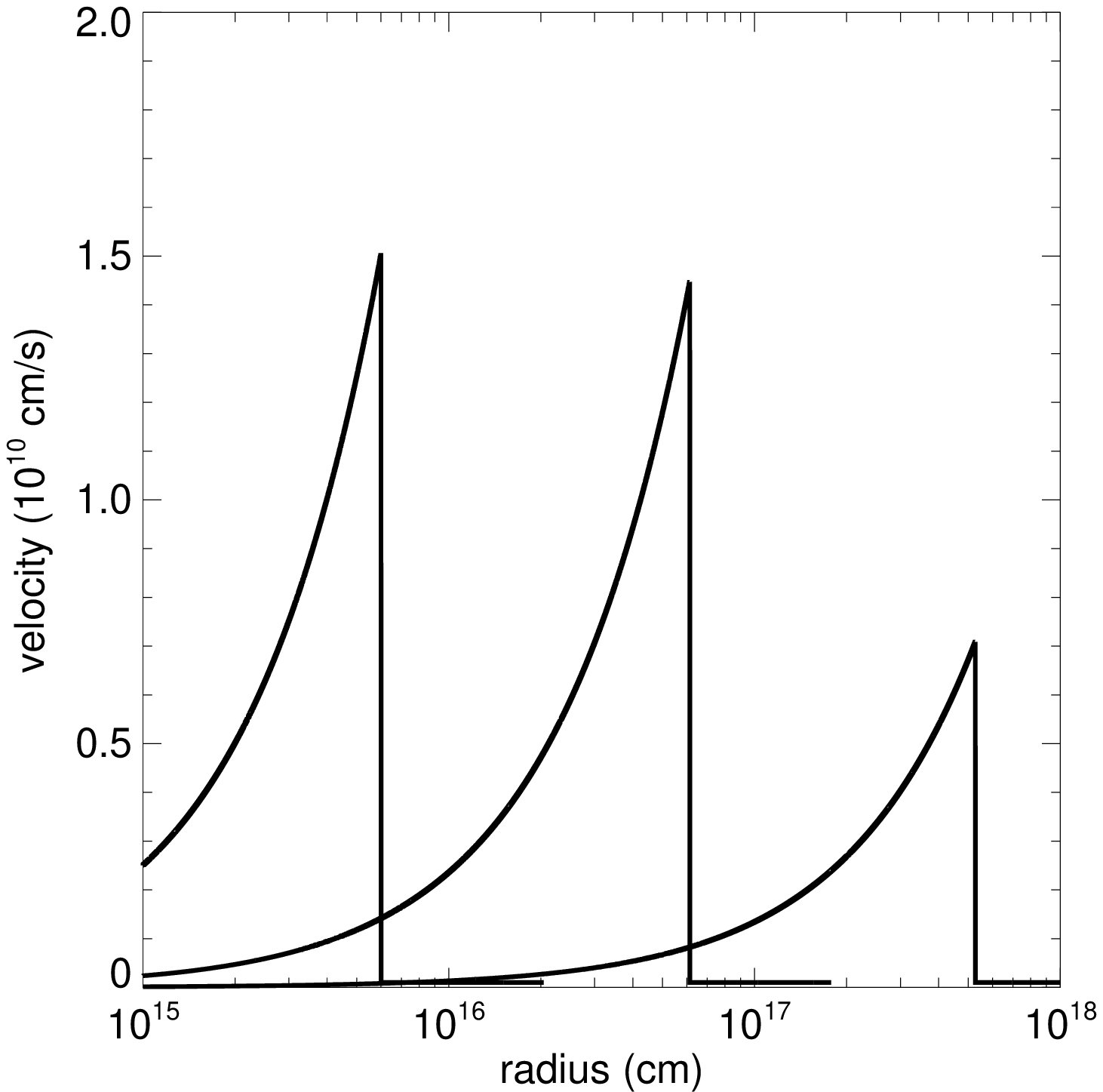,width=0.3\linewidth,clip=}  \\
\epsfig{file=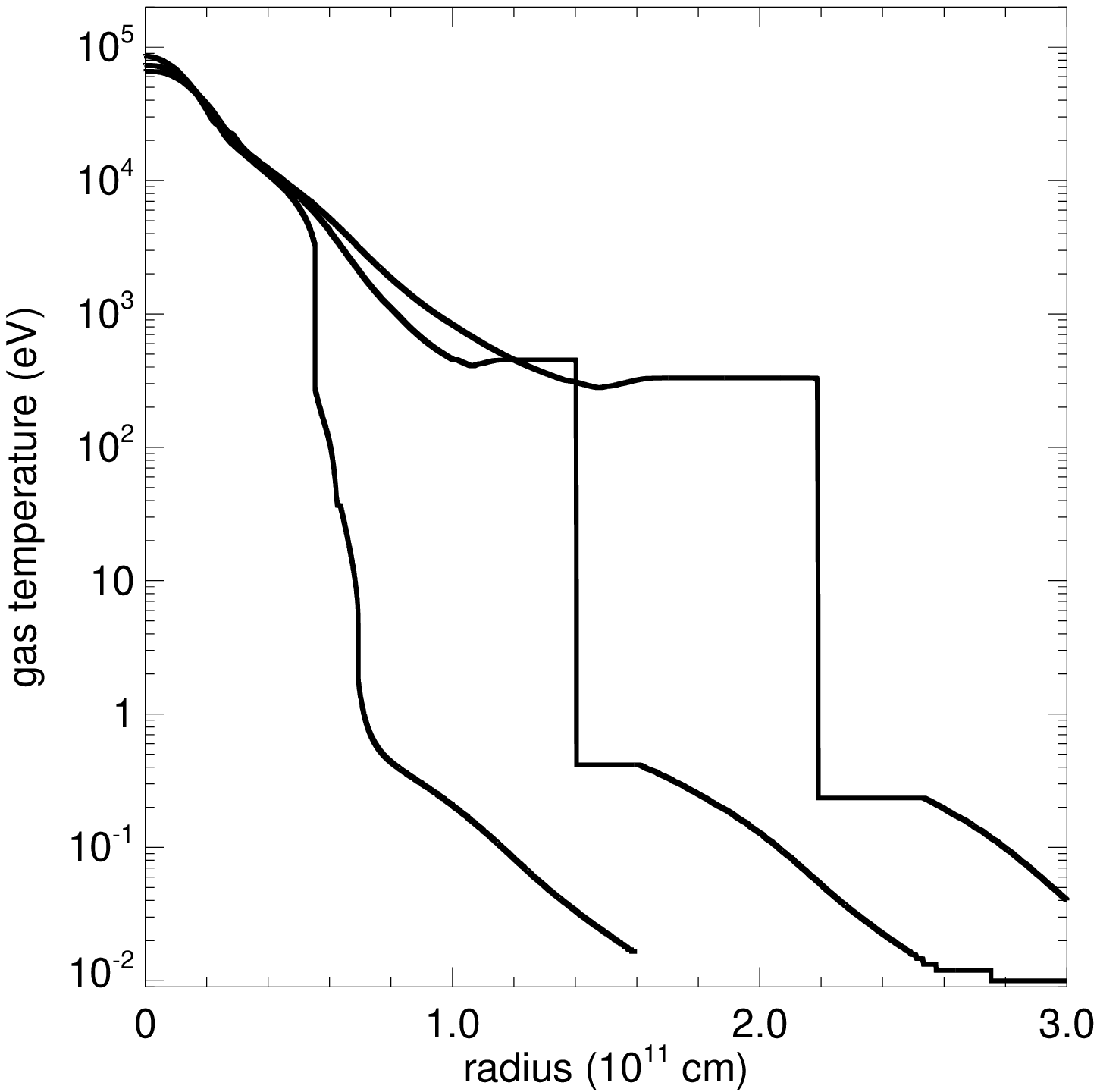,width=0.3\linewidth,clip=}  &
\epsfig{file=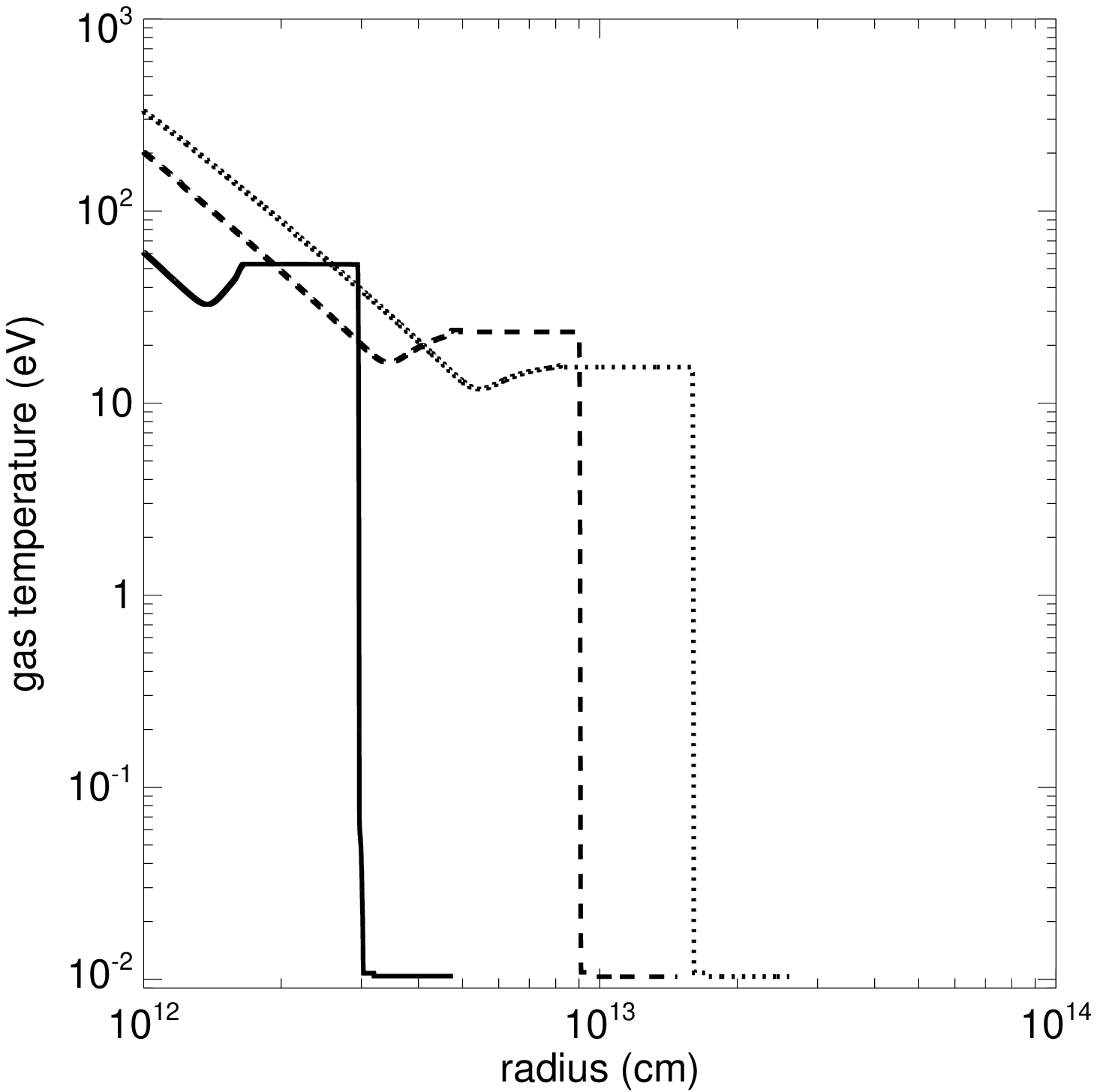,width=0.3\linewidth,clip=}  & 
\epsfig{file=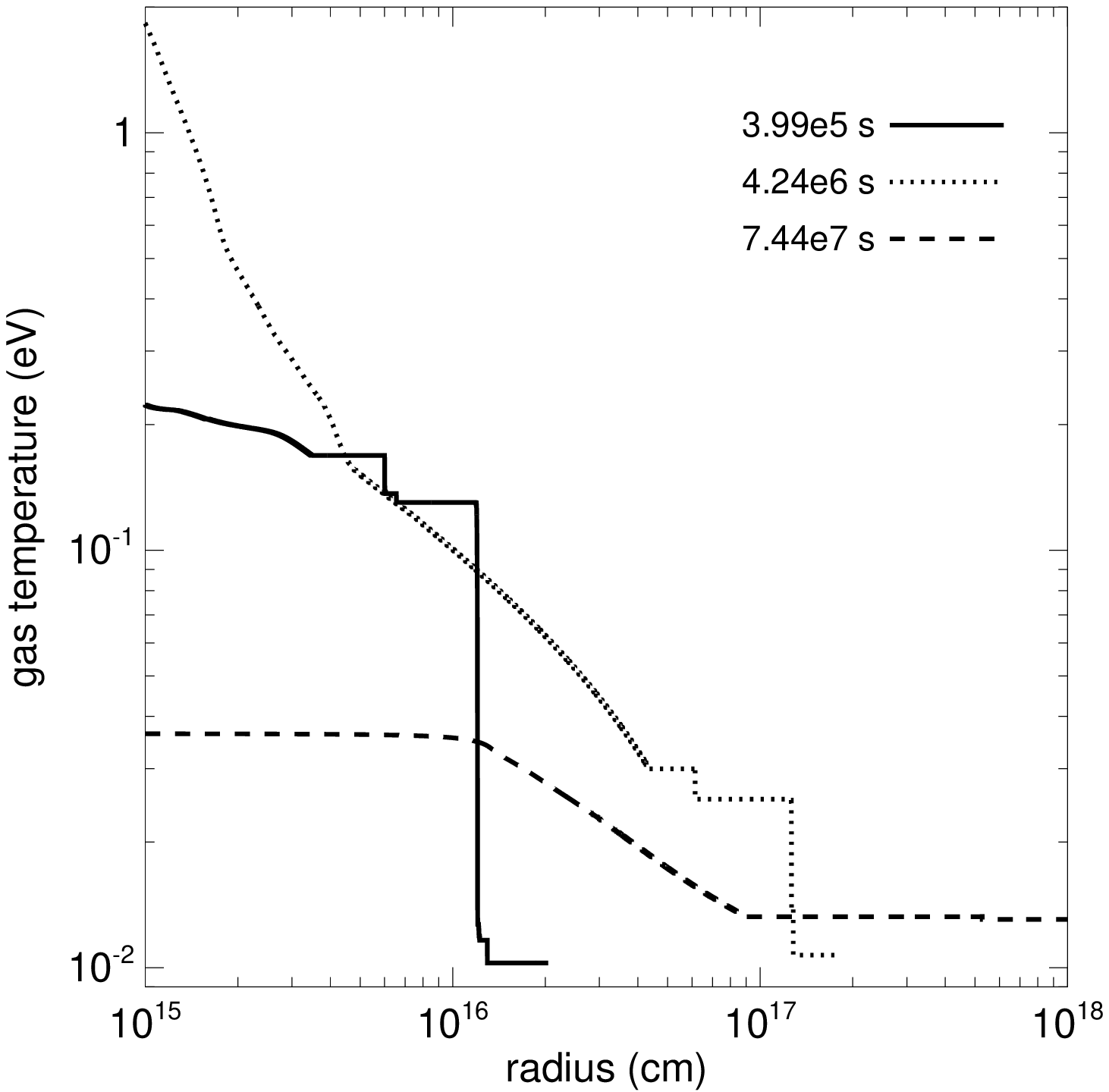,width=0.3\linewidth,clip=}  \\
\epsfig{file=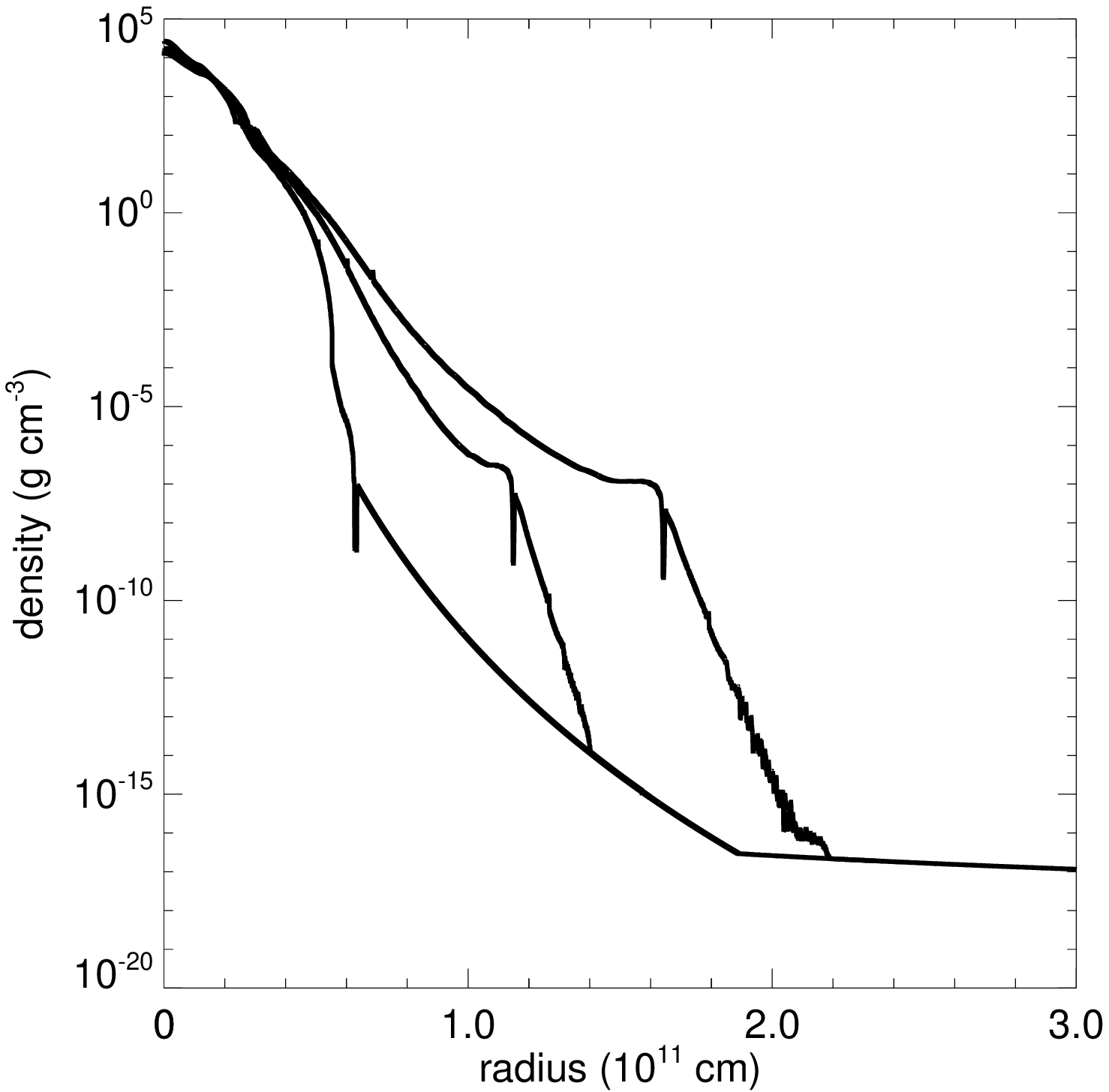,width=0.3\linewidth,clip=}  &
\epsfig{file=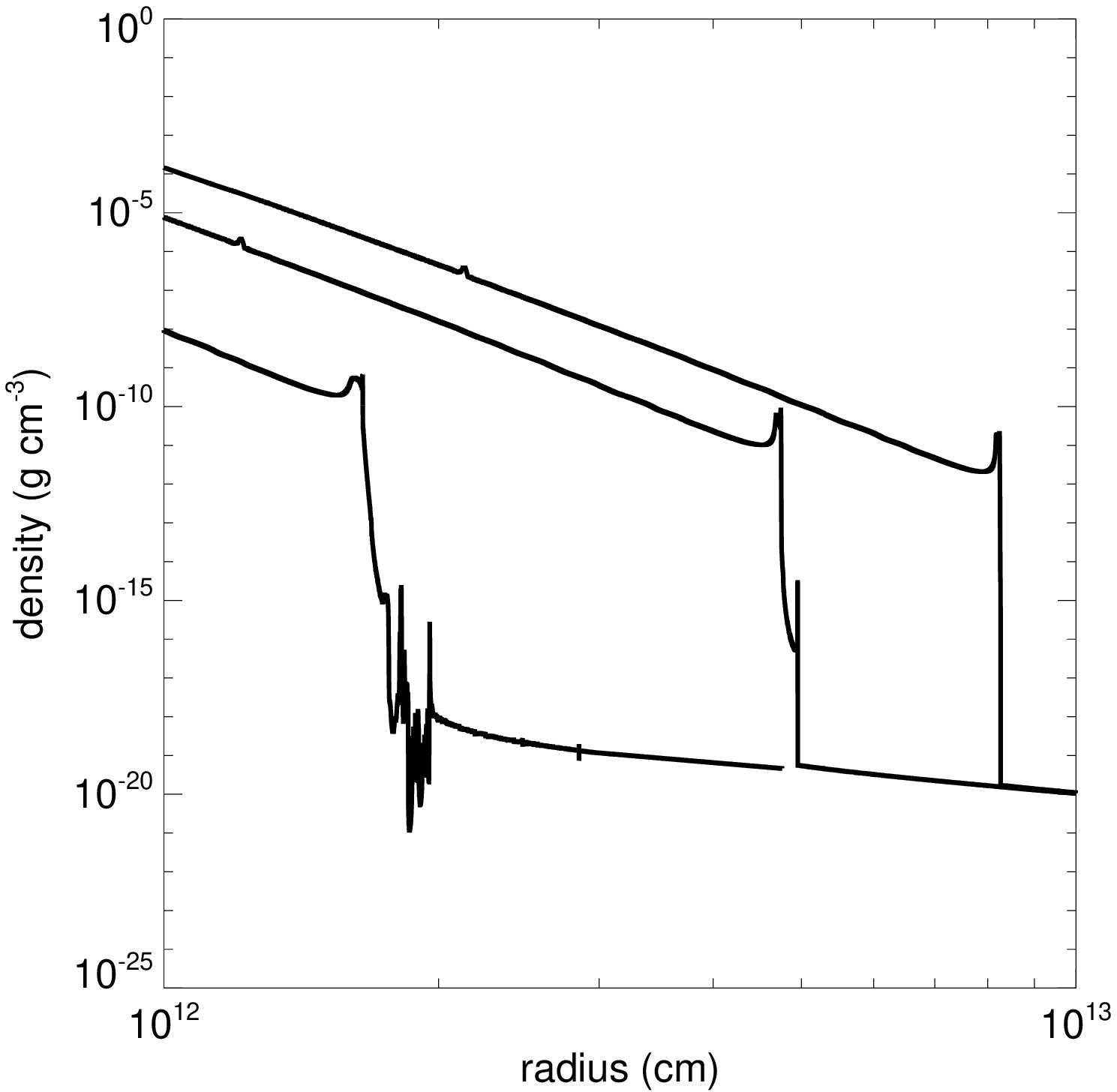,width=0.3\linewidth,clip=}  &
\epsfig{file=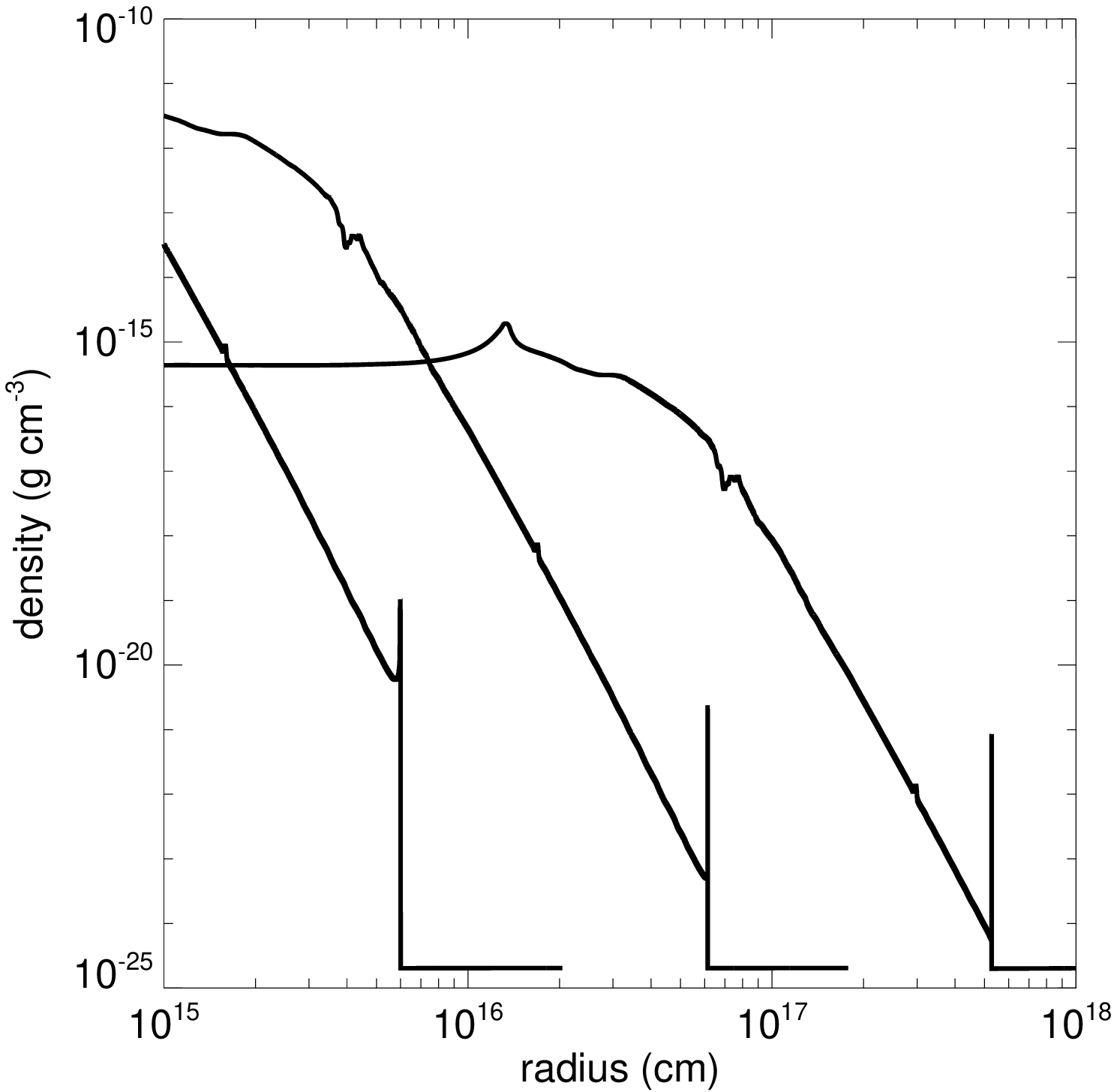,width=0.3\linewidth,clip=}
\end{tabular}
\end{center}
\caption{Hydrodynamic evolution of the a120 PI SN. Top: velocities; center: temperatures; 
bottom:  densities. Left: shock breakout. From left to right the times are 3.97 s, 7.17 s, and 
9.75 s. Center: intermediate evolution. From left to right, the times are 100 seconds, 302 
seconds and 533 seconds.  Right:  later evolution (the nebular phase).  From left to right, 
the times are 4.0e5 seconds, 4.24e6 seconds and 7.44e7 seconds.}
\label{fig:hydro}
\end{figure*}

We show density, temperature and velocity profiles for the a120 PI SN at shock breakout, 
at intermediate times, and at later times in Figure~\ref{fig:hydro}.  As it breaks out of the 
surface of the compact core and descends the density bridge, the shock accelerates to $
\sim$ 2.2 $\times$ 10$^{10}$ cm s$^{-1}$. As it approaches the bottom of the bridge, the 
shock begins to gradually slow down as it plows up the envelope. Within 1 - 2 seconds of 
breakout, photons that were previously advected along by the flow abruptly break free of 
the shock, as shown in the center left panel of Figure~\ref{fig:hydro}.  The breakout 
transient is visible as the flat plateau in gas energy ahead of the shock at 7.2 and 9.8 
seconds.  This radiation front initially heats the gas to $\sim$ 500 eV.  As the fireball 
expands, it cools by emitting radiation and performing work on the surrounding envelope.  
As it cools, its spectrum softens, and the temperature to which the radiation pulse heats 
the gas also decreases.

\begin{figure*}
\plottwo{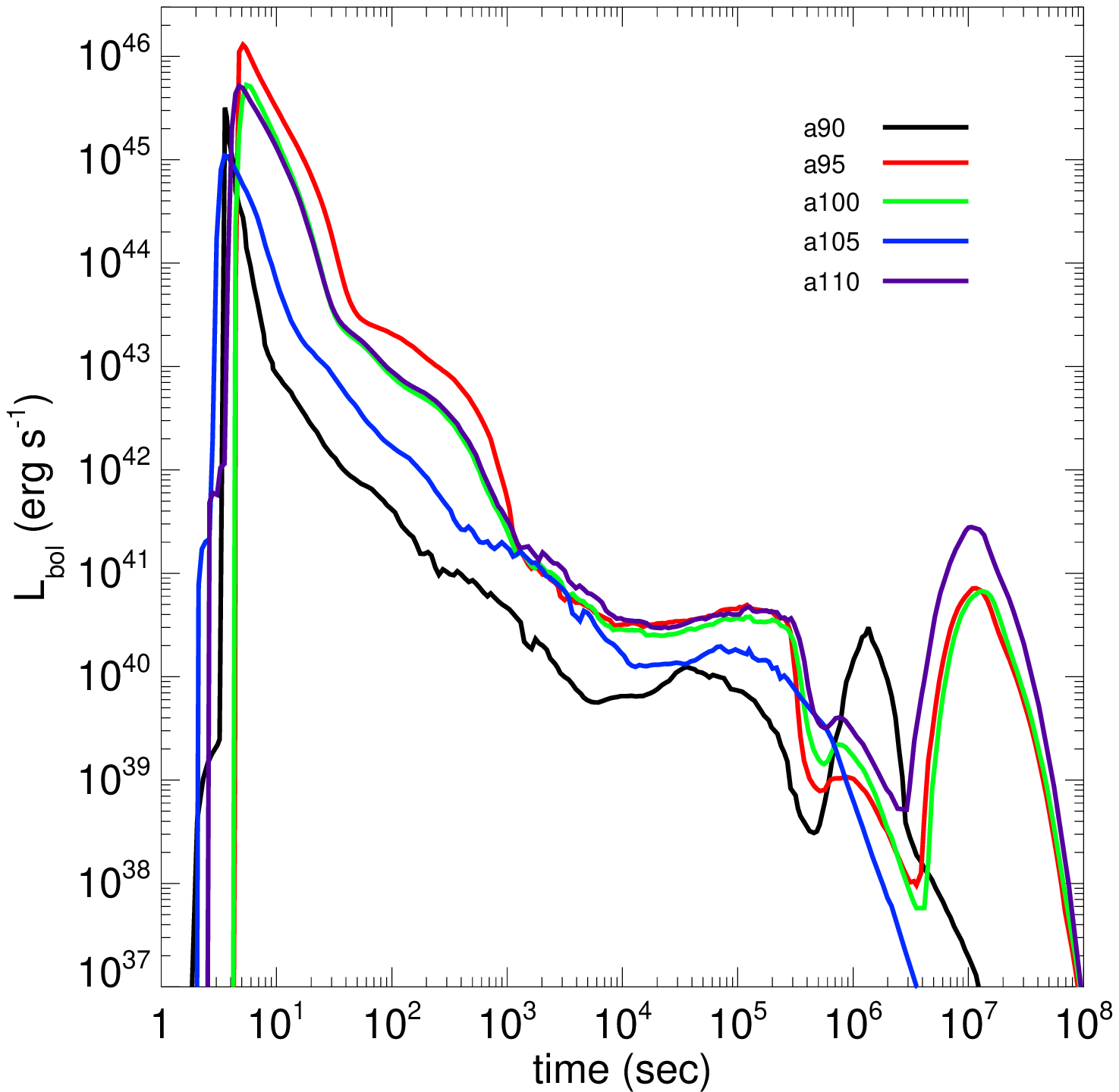}{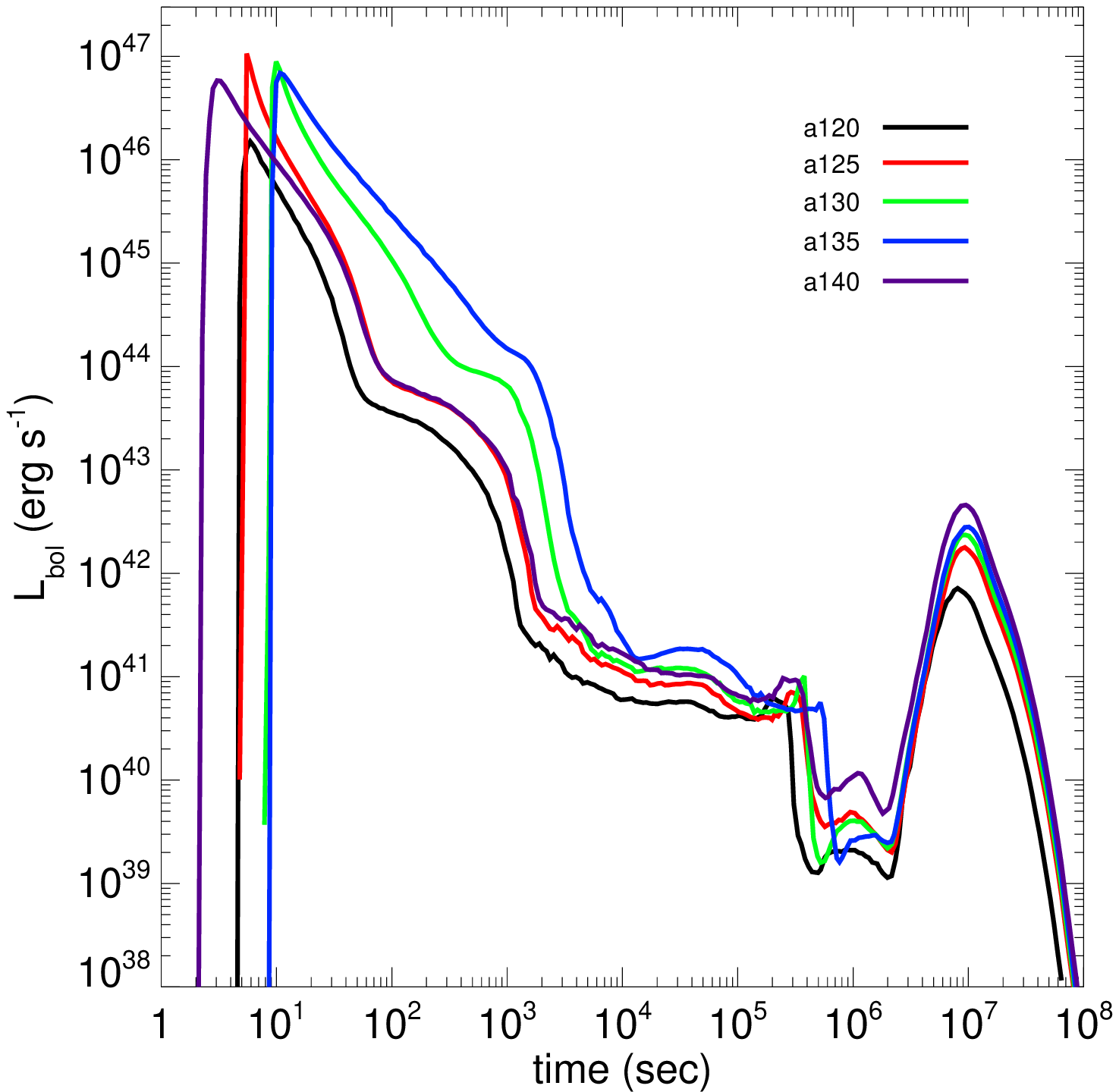}
\caption{Bolometric luminosities for 90 - 135 \Ms\ PI SNe.  Left panel:  90, 95, 100, 105 
and 110 \Ms\ SNe.  Right panel:  120, 125, 130, 135 and 140 \Ms\ explosions.} \vspace{
0.1in}
\label{fig:bLC}
\end{figure*} 

Bolometric luminosities for all 10 PI SNe in are shown in Figure~\ref{fig:bLC}. The duration 
of the breakout transient is greater than the light crossing time of the star, in part because 
the radiation remains partially coupled to the outer layers of the star that are blown off by 
the pulse.  As photons diffuse out through these outermost layers (the radiative precursor) 
they break free over a range of times and then become visible to an external observer. 
The opacity of the ejecta is also frequency dependent, so photons escape at different times 
according to their wavelengths \citep{bay14}.  As shown in Figure~\ref{fig:bLC}, breakout 
luminosities vary from $\sim$ 10$^{46}$ - 10$^{47}$ erg s$^{-1}$, and they generally rise 
with explosion energy.  Shock breakout also generally happens sooner in less massive stars 
because of their smaller radii.  The breakout pulse itself is composed mostly of X-rays and 
hard UV.  At $z \sim 20$ the pulse would last up to 1 - 2 days today, in principle making it 
much easier to detect at this epoch than in the local universe.  But although it is also the 
most luminous phase of the SN, shock breakout is least visible at high redshifts due to 
absorption by the neutral intergalactic medium (IGM).  Any X-rays that are not absorbed 
would redshifted into the far UV and absorbed in the outer layers of our Galaxy.
  
Radiation continues to drive the precursor ahead of the shock for $\sim$ 600 seconds, as 
shown in the center panels of Figure~\ref{fig:hydro}.  It is visible as the complex velocity 
and density structure at 2 $\times$ 10$^{12}$ cm at 302 s and 5 $\times$ 10$^{12}$ cm at 
533 s.  No strong reverse shocks form in the flow.  The shock eventually overtakes and 
merges with the precursor because as it expands and cools it dims, and its flux can no 
longer sustain it.  As shown in the panels on the right in Figure~\ref{fig:hydro}, the expansion 
of the flow is mostly homologous after 10$^5$ seconds.  All 10 PI SNe evolve through these 
stages in a similar manner.

\begin{figure}
\plotone{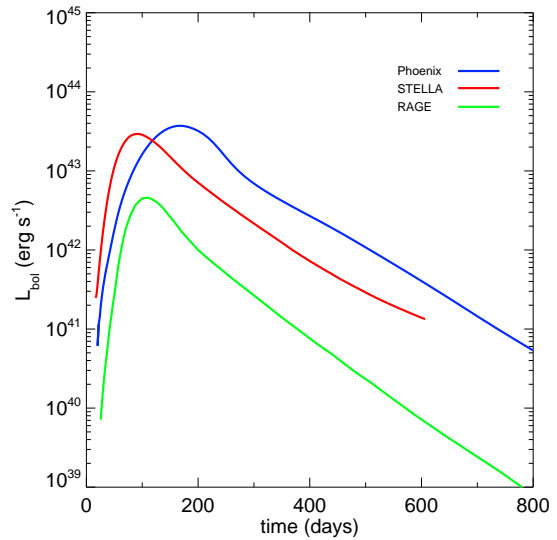}
\caption{Bolometric luminosities for the a140 PI SN calculated with RAGE, Phoenix and 
STELLA.} \vspace{
0.1in}
\label{fig:comp}
\end{figure}

At 10$^6$ - 10$^7$ s, the SNe rebrighten as photons from \Ni\ decay begin to diffuse out 
of the ejecta.  The range in peak times is due to the range in diffusion times from the \Ni\ 
layer to the surface for the progenitors in our study, with later times corresponding to larger 
ejecta masses.  Peak luminosities rise with \Ni\ mass, and the rebrightening typically lasts 
several hundred days in the rest frame of the SN.  RAGE predicts somewhat lower \Ni\ 
luminosities than SN codes that assume homologous expansion of the ejecta, as we show 
in Figure~\ref{fig:comp}.  Here, we plot bolometric luminosities for the a140 run calculated 
with the RAGE, Phoenix \citep{phoenix}, and STELLA \citep{stella} codes.  The Phoenix 
model is a 1D Lagrangian calculation with detailed S$_\mathrm{N}$ radiative transfer in a 
homologously expanding medium with 125 zones in mass.  The STELLA model is a 1D 
Lagrangian radiation hydrodynamics simulation with 100 energy groups and 125 zones in 
mass. 

As shown in Figure~\ref{fig:comp}, Phoenix predicts peak \Ni\ luminosities that are about 
an order of magnitude greater than those in RAGE.  This discrepancy is most likely due 
to two factors.  First, \Ni\ rebrightening begins at about 70 days after the SN, by which 
time 75\% of the total energy due to \Ni\ and \Co\ decay has been released ($\sim$ 1.3 
$\times$ 10$^{51}$ erg for the 7.3 \Ms\ of \Ni\ formed in the a140 explosion).  In RAGE, 
this energy is first deposited as heat locally in the ejecta and then transformed into kinetic 
energy as the hot \Co\ bubble performs $PdV$ work on its surroundings.  After the heat 
is transformed into kinetic energy it is difficult to recover it as luminosity later on when the 
\Ni\ layer is exposed to the IGM, unless the ejecta crashes into some kind of circumstellar 
structure.  This departure from the homologous expansion assumed in Phoenix is subtle 
because the total energy release due to radioactive decay is only 2\% of the kinetic energy 
of the ejecta, but it results in significantly less luminosity during rebrightening.  

The rebrightening in RAGE is therefore almost entirely due to \Co\ decay after 70 days, 
which is $\sim$ 25\% of the total decay energy.  On these numbers alone, one might 
expect the peak luminosity to be a factor of up to 4 lower in RAGE than in homologous 
expansion codes, which account for adiabatic expansion of the SN as a whole but do not 
capture the additional $PdV$ work done by the decay bubble.  The additional factor of 2 - 
3 less luminosity may be due to the lower density of the more expanded decay bubble in 
RAGE when it is exposed to the IGM.  We also note that both Lagrangian models may 
not have fully resolved the flow of radiation through the PI SN ejecta, allowing more of it 
to escape than really does.  When there are ten of thousands of optical depths in a given 
mesh point, numerical diffusion can allow photons to flow through the zone that should 
actually be absorbed.  This may partially explain the discrepancy between RAGE and 
STELLA, which is also a radiation hydrodynamical calculation that does not assume 
homologous expansion.  Opacities, minor differences in which can have substantial
effects on luminosities, may also contribute to the differences between these two codes.
More tests are now underway to study both effects on \Ni\ luminosity in a variety of SNe. 

\section{NIR Light Curves / Detection Limits}

Detections of SNe prior to the era of reionization ($z \gtrsim$ 6) require observations in 
the NIR because any flux blueward of the Lyman limit at higher redshifts is absorbed by 
the partially neutral IGM.  This likewise restricts detections in the optical to events at $z 
< $ 6.  All-sky surveys have the most potential to detect large numbers of high-$z$ SNe 
because their large survey areas can compensate for low star formation rates (SFRs) 
at early epochs \citep[e.g., Figure 3 of][]{wet13c}.  But extremely sensitive telescopes 
with more narrow fields such as {\it JWST}, the Thirty-Meter Telescope (TMT), the {\it 
Giant Magellan Telescope} ({\it GMT}) and the European Extremely Large Telescope 
(E-ELT) are still expected to detect appreciable numbers of Pop III SNe \citep{hum12}.  
We now consider detection limits in redshift for our PI SNe in the NIR for SNe at $z >$ 
6 and in the optical for events below this redshift. 
 
\begin{figure*}
\begin{center}
\begin{tabular}{c}
\epsfig{file=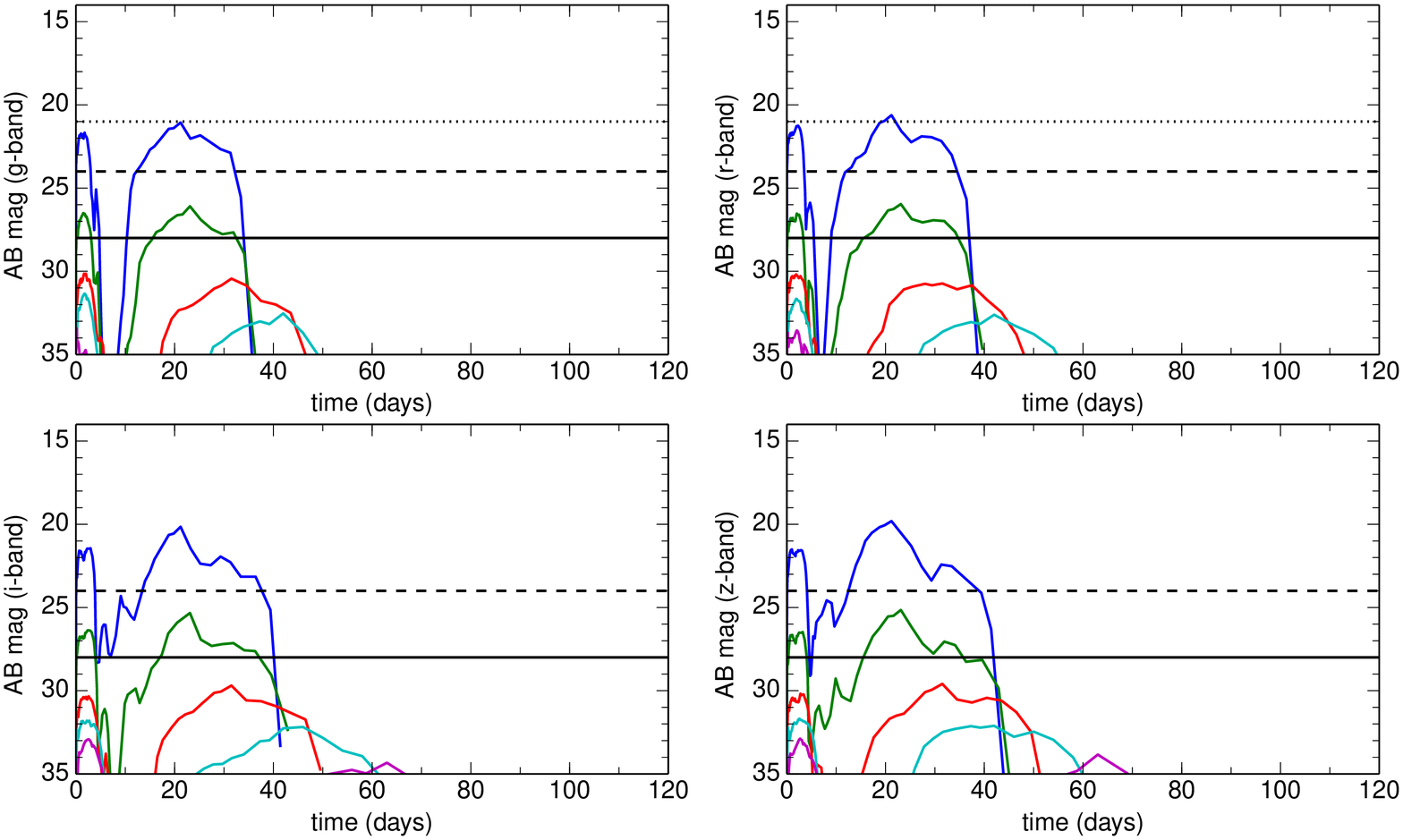,width=0.99\linewidth,clip=} \\
\epsfig{file=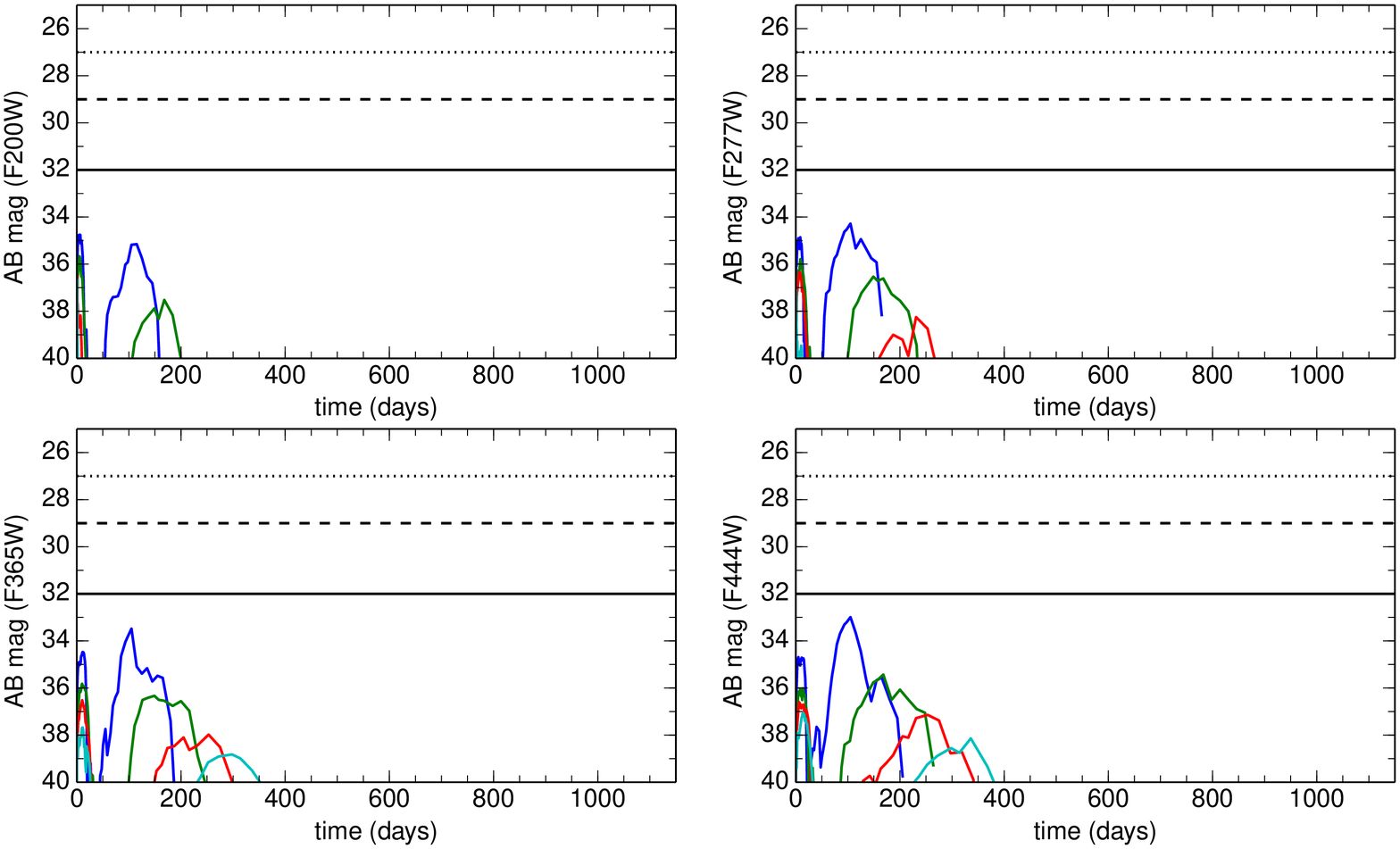,width=0.99\linewidth,clip=}
\end{tabular}
\end{center}
\caption{Light curves for the a90 PI SN at low redshifts (upper panels) and high redshifts 
(lower panels).  In the upper 4 panels, $z =$ 0.01 ({\it dark blue}), 0.1 ({\it green}), 0.5 ({\it 
red}), 1 ({\it light blue}), and 2 ({\it purple}).   The horizontal dotted, dashed and solid lines 
are photometry limits for PTF, Pan-STARRS and LSST, respectively.  In the lower 4 panels, 
$z =$ 4 ({\it dark blue}), 7 ({\it green}), 10 ({\it red}), 15 ({\it light blue}), 20 ({\it purple}) and 
30 ({\it yellow}).  The horizontal dotted, dashed and solid lines are photometry limits for 
WFIRST, WFIRST with spectrum stacking and {\it JWST}, respectively.}
\label{fig:a90}
\end{figure*}

\begin{figure*}
\begin{center}
\begin{tabular}{c}
\epsfig{file=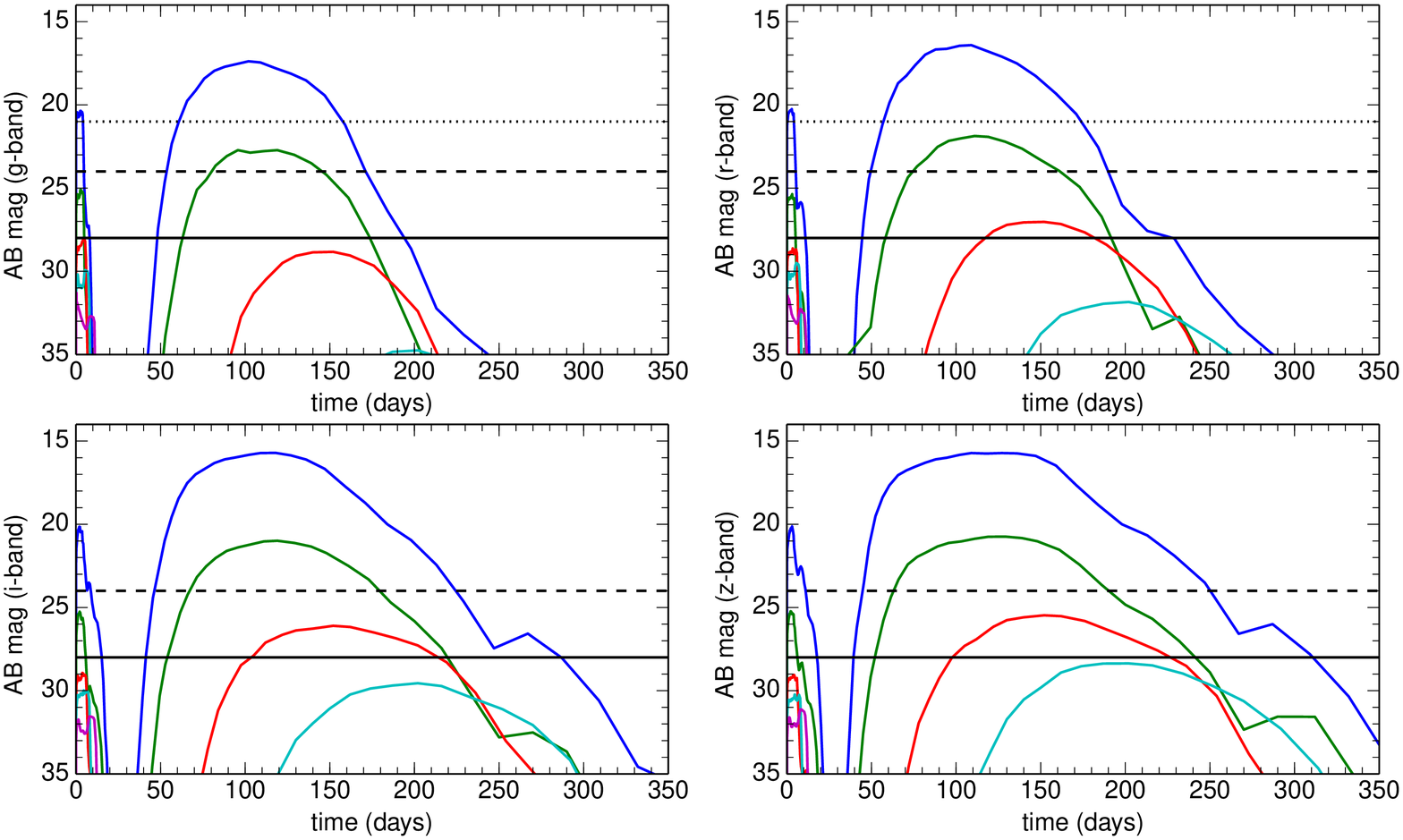,width=0.99\linewidth,clip=} \\
\epsfig{file=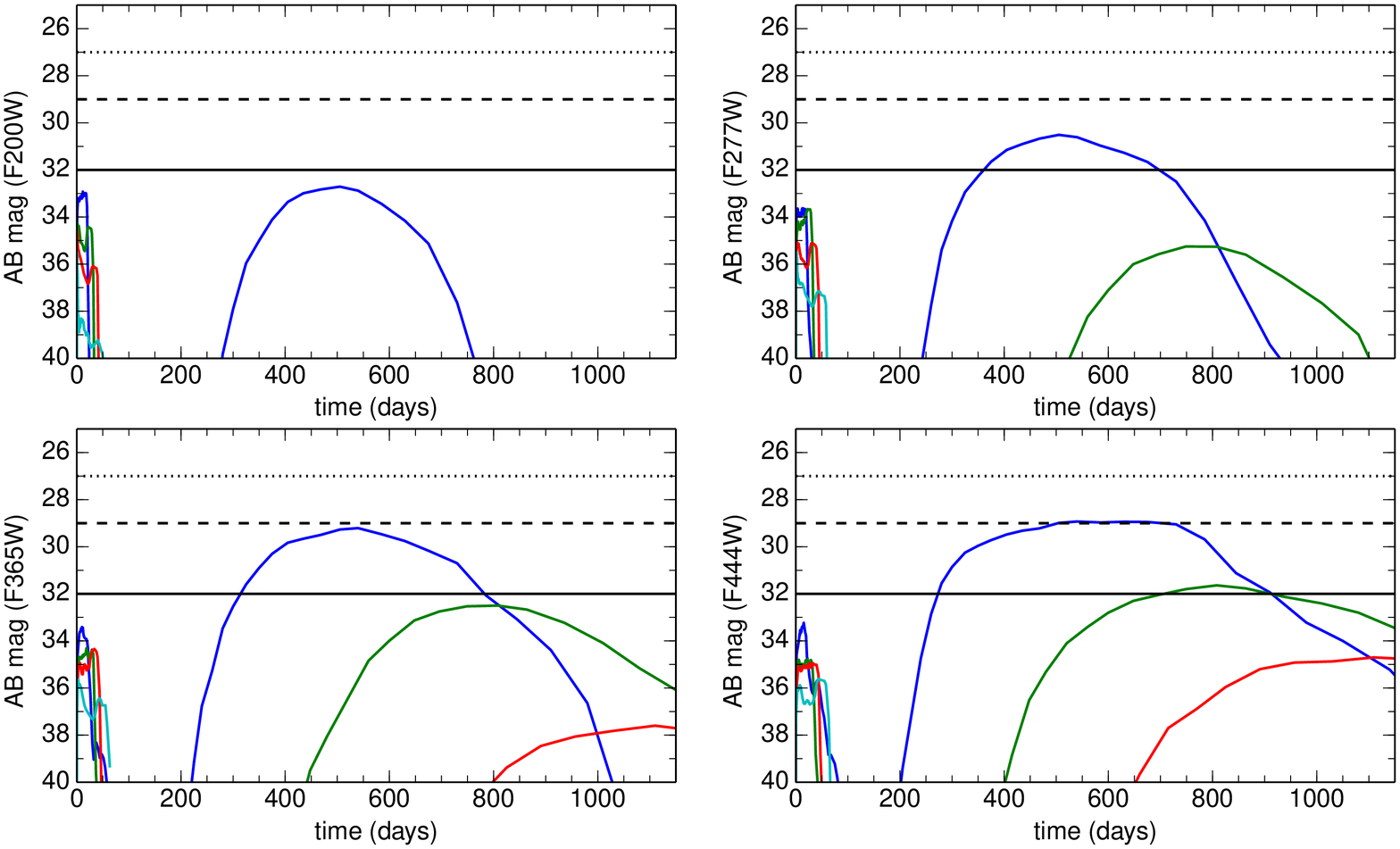,width=0.99\linewidth,clip=}
\end{tabular}
\end{center}
\caption{Light curves for the a120 PI SN at low redshifts (upper panels) and high redshifts 
(lower panels).  In the upper 4 panels, $z =$ 0.01 ({\it dark blue}), 0.1 ({\it green}), 0.5 ({\it 
red}), 1 ({\it light blue}), and 2 ({\it purple}).   The horizontal dotted, dashed and solid lines 
are photometry limits for PTF, Pan-STARRS and LSST, respectively.  In the lower 4 panels, 
$z =$ 4 ({\it dark blue}), 7 ({\it green}), 10 ({\it red}), 15 ({\it light blue}), 20 ({\it purple}) and 
30 ({\it yellow}).  The horizontal dotted, dashed and solid lines are photometry limits for 
WFIRST, WFIRST with spectrum stacking and {\it JWST}, respectively.}
\label{fig:a120}
\end{figure*}

\begin{figure*}
\begin{center}
\begin{tabular}{c}
\epsfig{file=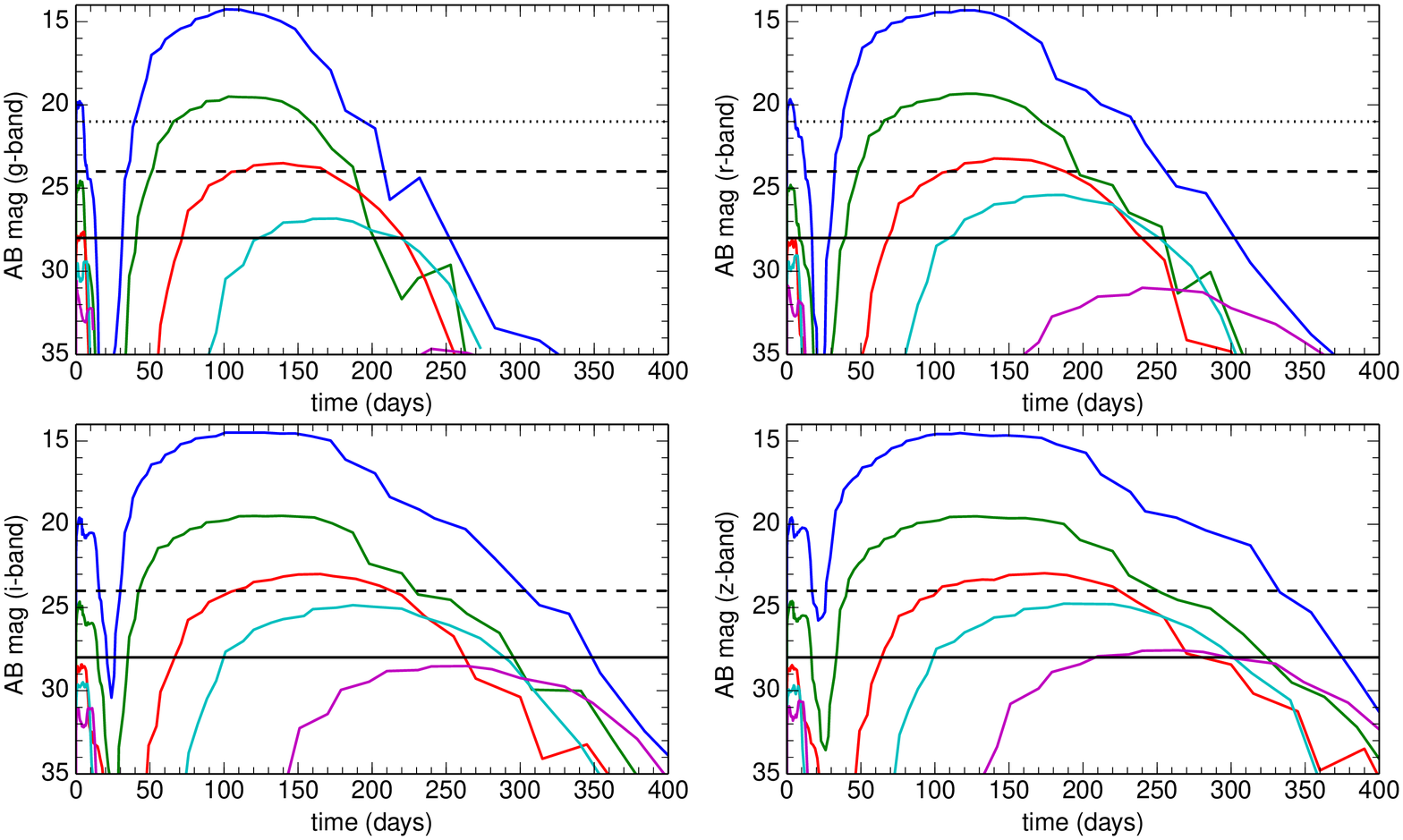,width=0.99\linewidth,clip=} \\
\epsfig{file=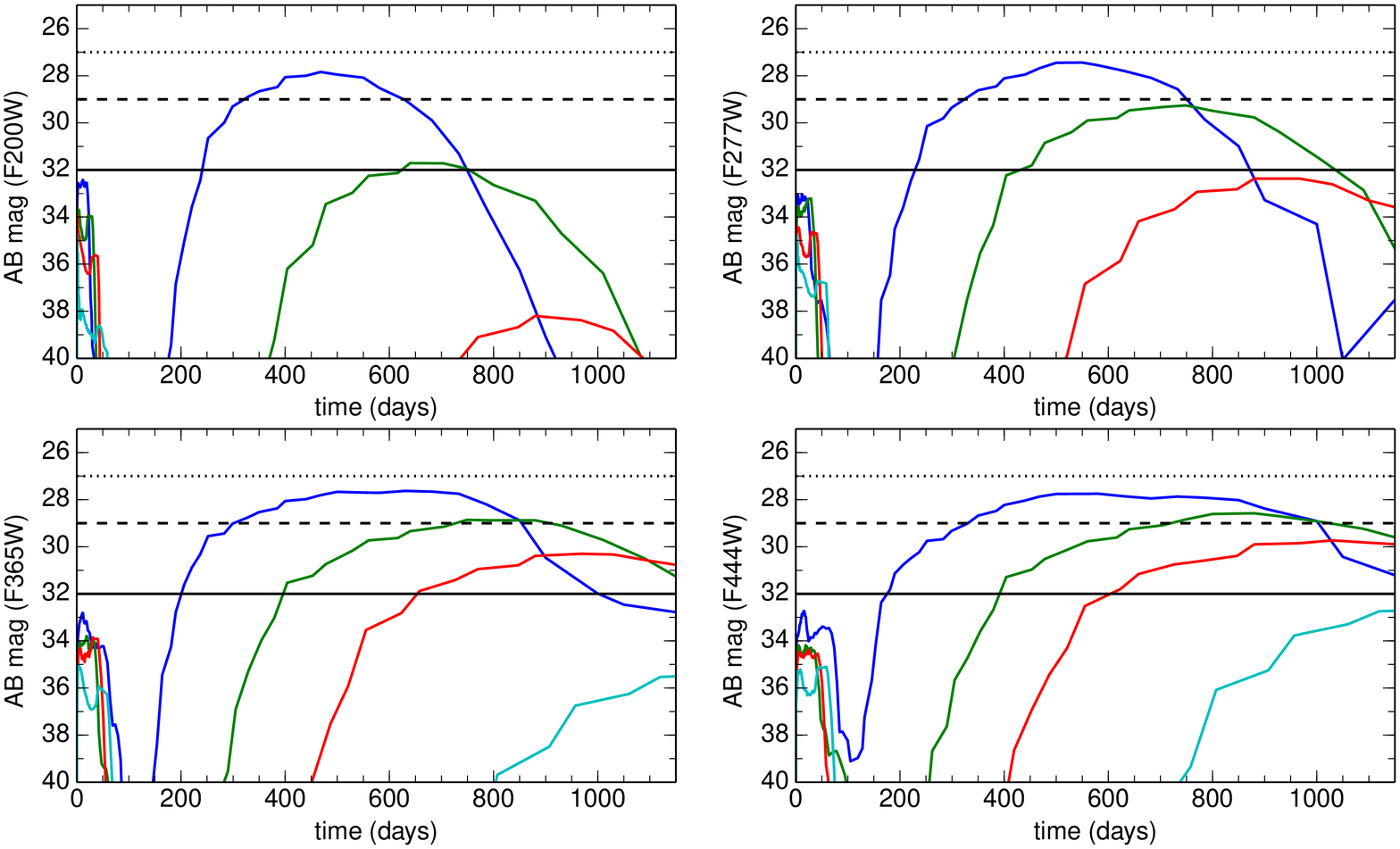,width=0.99\linewidth,clip=}
\end{tabular}
\end{center}
\caption{Light curves for the a140 PI SN at low redshifts (upper panels) and high redshifts 
(lower panels).  In the upper 4 panels, $z =$ 0.01 ({\it dark blue}), 0.1 ({\it green}), 0.5 ({\it 
red}), 1 ({\it light blue}), and 2 ({\it purple}).   The horizontal dotted, dashed and solid lines 
are photometry limits for PTF, Pan-STARRS and LSST, respectively.  In the lower 4 panels, 
$z =$ 4 ({\it dark blue}), 7 ({\it green}), 10 ({\it red}), 15 ({\it light blue}), 20 ({\it purple}) and 
30 ({\it yellow}).  The horizontal dotted, dashed and solid lines are photometry limits for 
WFIRST, WFIRST with spectrum stacking and {\it JWST}, respectively.}
\label{fig:a140}
\end{figure*}

We show optical and NIR light curves for the a90, a120 and a140 PI SNe in Figures 
\ref{fig:a90} - \ref{fig:a140} at low and high redshifts along with detection limits for {\it 
JWST}, WFIRST and the SN factories:  the Palomar Transient Factory (PTF), the 
Panoramic Survey Telescope \& Rapid Response System (Pan-STARRS) and the 
Large Synoptic Survey Telescope (LSST).  The light curves all have an initial short 
lived transient that lasts up to $\sim$ 50 days.  It is followed by a decline and then a 
second brighter and much longer lived phase that can last several hundred days or 
more depending on the filter.  This second broad peak is due to \Ni\ rebrightening.  
Detection limits in the NIR for these events vary widely with mass and explosion 
energy but range from a140 being visible to {\it JWST} out to $z \sim$ 7 - 10 for 500 
- 600 days to a90 only being visible at $z <$ 7.  Only the most energetic SNe are 
visible to WFIRST at $z \gtrsim$ 4, and only if their spectra can be stacked.

In the optical, detection limits in redshift vary from $z \sim$ 0.1 for the a90 PI SN 
to $z \sim$ 1 - 2 for a140.  In the g, r, i and z bands the light curves exhibit similar 
rise and fall times at a given redshift, but their durations increase with wavelength.  
They are visible in these bands for 50 - 250 days and exhibit enough variation to 
be recognized as transients, given the short cadences of the factories.  It is clear 
that the SN factories will not detect primordial PI SNe in this mass range, but {\it 
JWST} could find them out to $z \sim$ 7 - 10, when Pop III stars could still be 
forming in pockets of metal-free gas \citep{tss09,fop11}.  But determining whether 
or not such SNe are from Pop III stars would be problematic for several reasons.  
First, the high shock temperatures in these explosions would obscure the spectral 
lines that would differentiate these events from Pop II SNe.  There 
are also degeneracies in light curve structure between these SNe and the PI SNe 
of 0.1 - 0.3 \Zs\ stars studied by \citet{wet13e}.  By $z \sim$ 7 - 10 most stars are 
expected to be contaminated by metals from the first few generations of SNe in 
the universe.  Nevertheless, because these PI SNe can be easily distinguished 
from CC SNe they can be used to probe the masses of stars in the era of first 
galaxy formation and reionization.  

\section{Conclusion}

We find that 90 - 140 \Ms\ Pop III PI SNe whose progenitors have lost their H envelopes 
are only visible in the optical to PTF, Pan-STARRS and LSST out to $z \sim$ 1 - 2 but 
can be detected out to $z \sim$ 7 - 10 by {\it JWST} and the coming generation of 30 m 
telescopes.  These SNe fall into a now familiar pattern for highly energetic explosions of 
compact, massive Pop III stars that have shed their outer envelopes.  Although they 
exhibit very high luminosities and shock temperatures at breakout, 90 - 140 \Ms\ PI SNe, 
hypernovae \citep[HNe; ][]{smidt13a}, and the PI SNe studied by \citet{wet13e} are all 
much dimmer in the NIR at high redshift than 140 - 260 \Ms\ Pop III PI SNe with similar 
explosion energies, which can be detected at $z \gtrsim$ 30 \citep{wet12b}.  None of the 
compact core Pop III SNe in these three studies can be seen at $z \sim 15 - 20$, the era 
of the first stars.  Like the PI SNe considered here, HNe are only visible out to $z \sim$ 7 
- 10 to {\it JWST} and $z \sim$ 4 - 5 to WFIRST, with detections of 0.1 - 0.3 \Zs\ PI SNe 
by {\it JWST} being restricted to similar redshifts.  However, they could all easily appear in 
future surveys of the first galaxies, which will be principal targets of {\it JWST} and the 30 
m-class telescopes.

This picture could change if ejecta from the explosion crashes into the mass lost by the 
star prior to its death, which can result in a superluminous SN (SLSN) like SN 2006gy. 
These events can be far brighter in the NIR than the original explosion \citep{nsmith07b,
moriya10,chev11,moriya12}.  Their high luminosities are due to the large radius of the 
shell upon impact, 1 - 2 AU.  Much less energetic Type IIn SNe (1 - 2 foe) are visible to 
{\it JWST} at $z \sim$ 15 - 20 and to WFIRST at $z \sim$ 7 \citep{wet12e}, so it is quite
possible that the much more energetic collisions of the SNe in our study with shells may 
be visible to all-sky NIR missions out to $z \sim$ 10 - 15.  This would greatly increase 
their probability of detection at high $z$ because the wide survey areas of these 
missions could overcome low PI SN rates.  We are now simulating such explosions with 
RAGE.  

We have only considered PI SNe in very diffuse envelopes, in which all vestiges of the H 
layer have been driven beyond the immediate reach of the ejecta, as a first case.  How 
this gas is actually distributed in radius around the star when it dies depends on how its 
mass loss evolved over time, and the impact of such profiles on SN light curves has only 
begun to be studied. The large number of possibilities for PI SN progenitor structure, 
metallicity and envelope highlights the difficulty of matching any one PI SN candidate to 
current models.  Studies to date have only considered red supergiants, blue compact 
giants, and stripped He cores.  Stars of intermediate radius, such as yellow supergiants 
(YSGs), are only now being studied \citep[see][who have found that the PI SN of a 250 
\Ms\ YSG yields a bolometric light curve that is a good fit to SN 2007bi]{kz14a}. 

Although these less massive PI SNe will not be visible among the first generation of 
stars, they can be used to probe the stellar populations of the first galaxies and cosmic 
SFRs in the era of cosmological reionization.  They, together with a growing number of 
other types of SNe, will soon open a direct window on star formation in the primeval 
universe.

\acknowledgments

DJW acknowledges support from the European Research Council under the European 
Community's Seventh Framework Programme (FP7/2007-2013) via the ERC Advanced 
Grant "STARLIGHT: Formation of the First Stars" (project number 339177).  EC would like 
to thank the Enrico Fermi Institute for its support via the Enrico Fermi Fellowship.  JS was 
supported by a LANL LDRD Director's Fellowship. Work at LANL was done under the 
auspices of the National Nuclear Security Administration of the U.S. Department of Energy 
at Los Alamos National Laboratory under Contract No. DE-AC52-06NA25396. All RAGE and 
SPECTRUM calculations were performed on Institutional Computing (IC) platforms at LANL 
(Pinto, Lobo, Moonlight and Wolf).

\bibliographystyle{apj}
\bibliography{refs}

\end{document}